\documentclass[prd,prd,10pt,twocolumn,superscriptaddress,nofootinbib ,noshowpacs,preprintnumbers]{revtex4-1}
\usepackage{xcolor}
\usepackage{graphicx}
\usepackage{dcolumn}
\usepackage{bm}
\usepackage{amsmath}
\usepackage{amssymb}
\usepackage[colorlinks]{hyperref}
\usepackage{multirow}
\usepackage{wasysym}

\newcommand\Tstrut{\rule{0pt}{2.6ex}}  
\newcommand\Bstrut{\rule[-1.2ex]{0pt}{0pt}}

\begin{document}


\title{RES-NOVA: A new neutrino observatory based on archaeological lead}

\author{Luca Pattavina}
 \email{luca.pattavina@lngs.infn.it}

\affiliation{%
 Physik-Department, Technische Universit{\"a}t M{\"u}nchen, 85747 Garching, Germany
}%
 \affiliation{%
INFN, Laboratori Nazionali del Gran Sasso, 67100 Assergi, Italy
}
 
\author{Nahuel Ferreiro Iachellini}%
 \email{ferreiro@mpp.mpg.de}
\affiliation{%
 Max-Planck-Institut f{\"u}r Physik, 80805 M{\"u}nchen, Germany
}%

\author{Irene Tamborra}
 \email{tamborra@nbi.ku.dk}
\affiliation{Niels Bohr International Academy and DARK, Niels Bohr Institute, University of Copenhagen, Blegdamsvej 17, 2100, Copenhagen, Denmark}%

\date{\today}

\begin{abstract}
We propose the RES-NOVA project which will hunt neutrinos from core-collapse supernovae (SN)  via coherent elastic neutrino-nucleus scattering (CE$\nu$NS) using an array of archaeological lead (Pb) based cryogenic detectors. The high CE$\nu$NS cross-section on Pb and the ultra-high radiopurity of archaeological Pb enable the operation of a high statistics experiment equally sensitive to all neutrino flavors with reduced detector dimensions in comparison with existing Neutrino Observatories, and easy scalability to larger detector volumes. RES-NOVA  is planned to  operate according to three phases with increasing detector volumes:  (60~cm)$^3$, (140~cm)$^3$, and ultimately 15$\times$(140~cm)$^3$. It will be sensitive to SN bursts  up to Andromeda with 5$\sigma$ sensitivity with already existing technologies and will have excellent energy resolution with $1$~keV threshold. Within our Galaxy, it will be possible to discriminate core-collapse SNe from black hole forming collapses with no ambiguity even in the first phase of RES-NOVA. The average neutrino energy of all flavors, the SN neutrino light curve, and the total  energy emitted in neutrinos  can potentially be constrained with a precision of few $\%$ in the final detector phase. RES-NOVA will be sensitive to flavor-blind neutrinos from the diffuse SN neutrino background with an exposure of $620$~ton $\cdot$ y. The proposed RES-NOVA project has the potential to lay down the foundations for a new generation of neutrino telescopes, while relying on a very simple technological setup.

\end{abstract}


\maketitle


\section{\label{sec:level1}Introduction}

Stars heavier than $8\ M_\odot$ end their life giving birth to some of the most energetic transients in our Universe: core-collapse supernovae (SN)~\cite{Janka:2006fh,Burrows:2012ew}. Neutrinos play a major role in the SN mechanism~\cite{Janka:2016fox}. During the SN explosion, $10^{58}$ neutrinos are emitted; the detection of SN neutrinos would be enormously important to disclose  information on the physics of the core collapse, not otherwise accessible~\cite{Muller:2019upo,Mirizzi:2015eza,Horiuchi:2017sku}. Despite the steep progress in the field, many questions revolving around  the  SN mechanism remain unanswered.  
The detection of neutrinos from a SN burst occurring within our galaxy will also shed light on the yet poorly understood behavior of neutrinos at extreme densities~\cite{Mirizzi:2015eza,Shalgar:2019qwg,Horiuchi:2017sku} as well as on any non-standard neutrino properties~\cite{Mirizzi:2015eza,Shalgar:2019rqe,Suliga:2019bsq,Mastrototaro:2019vug,Sung:2019xie,Wu:2013gxa,deGouvea:2019goq,Stapleford:2016jgz,EstebanPretel:2007yu}.  To maximize the amount of information that can be extracted from the neutrino signal and precisely reconstruct the neutrino emission properties, it will be crucial to detect all six neutrino flavors~\cite{Drukier:1983gj,Beacom:2002hs,Horowitz:2003cz}.

Existing neutrino detectors with large target mass [$\mathcal{O}(10~\mathrm{kton})$] are sensitive to SN neutrinos, in addition to others being planned or under construction~\cite{Mirizzi:2015eza,Scholberg:2017czd}. Astrophysical neutrinos can be detected via weak charge-current (CC) and neutral-current (NC) interactions on protons and electrons. The most significant interaction processes are inverse-beta decay (IBD) and elastic scattering on electrons (ES). 

The IBD requires target material elements such as water or organic scintillators, where a large number of free protons is available. This channel is sensitive to $\overline\nu_e$, and some experiments can detect the positron annihilation. A measurement of the positron energy can lead to a precise estimation of the SN neutrino energy. This is the case of  KamLAND~\cite{Asakura:2015bga}, Borexino~\cite{Cadonati:2000kq} and LVD~\cite{Agafonova:2006fz}  that are using large volume organic liquid scintillators. In addition,  JUNO~\cite{An_2016}, which is under commissioning,  will be online within the next few years from now with an extremely large detector mass of $20$~kton.

Super-Kamiokande~\cite{Vagins:2019yls} runs a large volume water Cherenkov detector of 32 kton. Thanks to its recent upgrade, where gadolinium was dissolved into water, the neutrino interaction signal (IBD of $\overline{\nu}_e$) can be tagged with a higher efficiency ($90\%$~\cite{GalloRosso:2017mdz}) via neutron capture on gadolinium. This additional feature has the potential to strongly improve the  sensitivity of Super-Kamiokande to SN neutrino interactions, especially for what concerns the detection of the diffuse background of SN neutrinos (DSNB). The IceCube Neutrino Telescope~\cite{Abbasi:2011ss} is a Cherenkov detector in ice with the potential to provide the largest statistics of SN neutrino events. On the other hand, the upcoming liquid argon facility DUNE will allow to detect  $\nu_e$'s with higher statistics~\cite{Migenda:2018ljh}.

The ES is also a relevant detection channel for SN neutrinos, being  sensitive to all neutrino flavors through NC interactions (while the CC channel  is exploitable for $\overline\nu_e$ and $\nu_e$ as well). Unfortunately, in conventional detectors the ES has an interaction cross-section which is some orders of magnitude lower than IBD and it can contribute for a few percents to the total SN event rate~\cite{Scholberg:2012id}.
Other detection channels may also be used, such as CC interactions on nuclei~\cite{Drukier:1983gj} or protons~\cite{Beacom:2002hs,Dasgupta:2011wg}. 
 However, these processes -- especially the interactions on nuclei -- have much lower cross-sections. Furthermore, they are affected by large theoretical uncertainties and lack of experimental data at the energy scale of interest. This is the case of lead or iron targets proposed for SN neutrino detection via inelastic neutrino-nucleus CC interactions~\cite{Kolbe:2000np,Volpe:2001gy,Vaananen:2011bf}.

In this picture, the available experimental approaches lack of a  detection channel equally and highly sensitive to all neutrino flavors.
The coherent elastic neutrino-nucleus scattering (CE$\nu$NS) is an especially intriguing option thanks to its sensitivity to all neutrino flavors (NC process) and its high interaction cross section. The latter  scales as the square of the neutron number of the target nucleus at low energies~\cite{Freedman:1977xn,Akimov:2017ade}.

In the context of SN neutrino detection, CE$\nu$NS detectors would allow an estimation of the overall neutrino emission properties without the uncertainties related to standard oscillation physics~\cite{Mirizzi:2015eza}. In addition, CE$\nu$NS-based detectors complement dedicated neutrino telescopes, allowing to potentially improve our chances to reconstruct the  emission properties of the non-electron (anti)neutrinos. Direct detection dark matter experiments exploit CE$\nu$NS~\cite{Undagoitia:2015gya,Raj:2019wpy} and can also act as neutrino telescopes, if the target material is larger than a few tons~\cite{Lang:2016zhv,Newstead:2020fie,Horowitz:2003cz,Monroe:2007xp,Strigari:2009bq,XMASS:2016cmy}. As pointed out in \cite{Raj:2019sci}, they are able to identify different characteristics of the SN event.

In this paper, we propose an innovative approach for the detection of SN neutrinos: the archaeological lead (Pb) cryogenic detector RES-NOVA. The RES-NOVA project aims to exploit CE$\nu$NS for the detection of astrophysical neutrinos. Lead is the only element of the periodic table that simultaneously offers the highest CE$\nu$NS cross-section, for a high interaction rate, and the highest nuclear stability, for ultra-low background levels. Already in its smallest size, the proposed detector would allow to obtain competitive high statistics of neutrino events through CE$\nu$NS by employing a detector with a miniaturized volume, compared to the currently running neutrino observatories. Furthermore, the employment of archaeological Pb will secure a background level which will be some orders of magnitude lower than other Pb samples, thus enabling a high statistical significance detection of SN neutrinos. 
The high CE$\nu$NS cross section of Pb, the ultra-low background of archaeological Pb and the cutting-edge performance of solid state cryogenic detectors, will allow the proposed experiment to reach out to SNe up to $\mathcal{O}(1~\mathrm{Mpc})$ and with the potential to reconstruct the neutrino average energy with an uncertainty of few $\%$.

The structure of the paper is as follows. Section~\ref{sec:nusignal} provides a brief introduction on the neutrino emission properties from the six benchmark SN models adopted in this work. We discuss the potential of exploiting CE$\nu$NS in a lead-based cryogenic detector, introduce the main characteristics of our proposed technology, and discuss the expected backgrounds and the detector energy response in Sec.~\ref{sec:CEnuNS}. The event rate for a Galactic SN is estimated in Sec.~\ref{sec:SNburst}; we also explore the possibility of distinguishing among different SN models and nuclear EoS, and reconstruct the average energy of the SN neutrinos. The possibility of detecting the DSNB with RES-NOVA is explored in Sec.~\ref{sec:dsnb}. Finally, conclusions are presented in Sec.~\ref{sec:conclusions}.

\section{Neutrino emission from core-collapse supernovae and black hole forming collapses}\label{sec:nusignal}

For a SN at distance $d$ from the Earth (assumed to be $d \simeq 10$~kpc for a Galactic burst), the differential flux for each neutrino  $\nu_\beta$  ($\nu_e$, $\bar\nu_e$, and $\nu_x=\nu_{\mu,\tau},\bar\nu_{\mu,\tau}$) at the time $t$ after the SN core bounce is 
\begin{equation}
\label{eq:nuflux} 
f_{\beta}^0(E,t)=\frac{L_{\beta}(t)}{4 \pi d^2} \frac{\phi_\beta(E,t)}{\langle E_\beta(t) \rangle}\ ,    
\end{equation}
with $L_{\beta}(t)$ being the time-dependent neutrino luminosity and $\langle E_\beta(t) \rangle$ the $\nu_\beta$ mean energy. The neutrino energy distribution $\phi_\beta(E,t)$ 
is parametrized as~\cite{Tamborra:2012ac,Keil:2002in}:
\begin{equation}
\phi_\beta(E,t)=\xi_\beta(t) \left(\frac{E}{\langle E_\beta(t)\rangle}\right)^{\alpha_\beta(t)} \exp\left(-\frac{(\alpha_\beta(t)+1) E}{\langle E_\beta(t)\rangle}\right)\ ,
\end{equation}
where  $\alpha_\beta(t)$ is such that
\begin{equation}
\frac{\langle E_\beta(t)^2\rangle}{\langle E_\beta(t)\rangle^2} = \frac{2+\alpha_\beta(t)}{1+\alpha_\beta(t)}\ ,    
\end{equation}
and $\xi_\beta(t)$ is obtained by $\int dE\ \phi_\beta(E,t)=1$. 

In order to estimate the event rate expected in RES-NOVA, we rely on the neutrino emission properties from a set of one-dimensional (1D) spherically symmetric hydrodynamical SN simulations by the Garching group~\cite{garch,Mirizzi:2015eza}. In order to take into account the variability of the expected neutrino signal as a function of the SN evolution outcome (i.e.~standard core-collapse SNe and black hole forming stellar collapses also dubbed failed SNe in the following), SN mass, and nuclear equation of state (EoS), we adopt six different SN models. Four models of standard core-collapse SNe with mass of $9.6$ and $27\ M_\odot$, each model with Lattimer and Swesty EoS~\cite{Lattimer:1991nc} with nuclear incompressibility modulus $K=220$~MeV (LS220 EoS) and with the SFHo hadronic EoS (SFHo EoS)~\cite{Steiner:2012rk}. The remaining two models are  black hole forming collapses with a mass of $40\ M_\odot$: s40c and s40s7b2c, corresponding to ``slow'' and ``fast'' black hole formation (i.e.~with different mass accretion rate)~\cite{Mirizzi:2015eza}. Both  black hole forming models have been simulated with LS220 EoS. 

\begin{figure*}[t]
\centering
\includegraphics[width=0.49\textwidth]{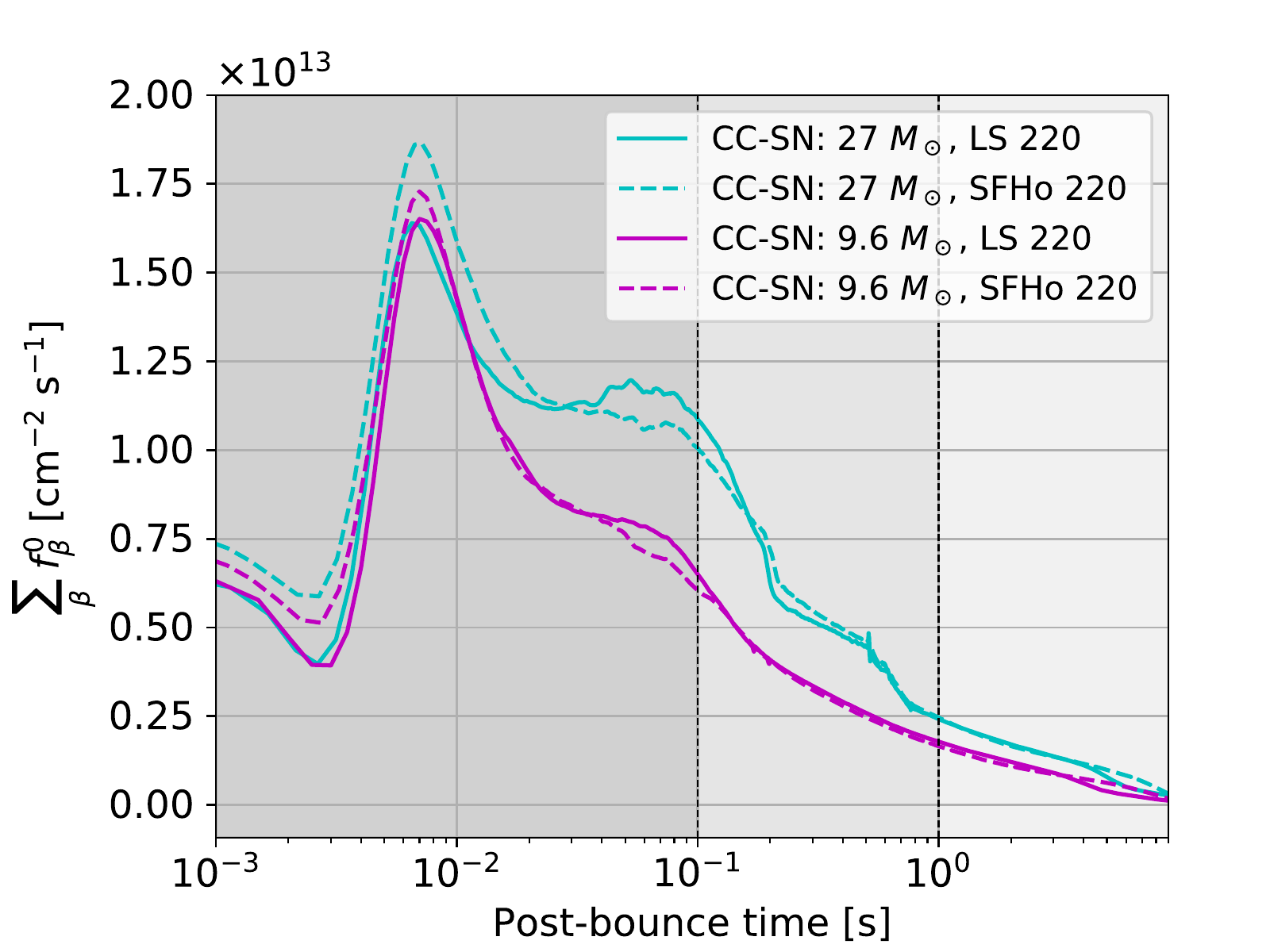}
\includegraphics[width=0.47\textwidth]{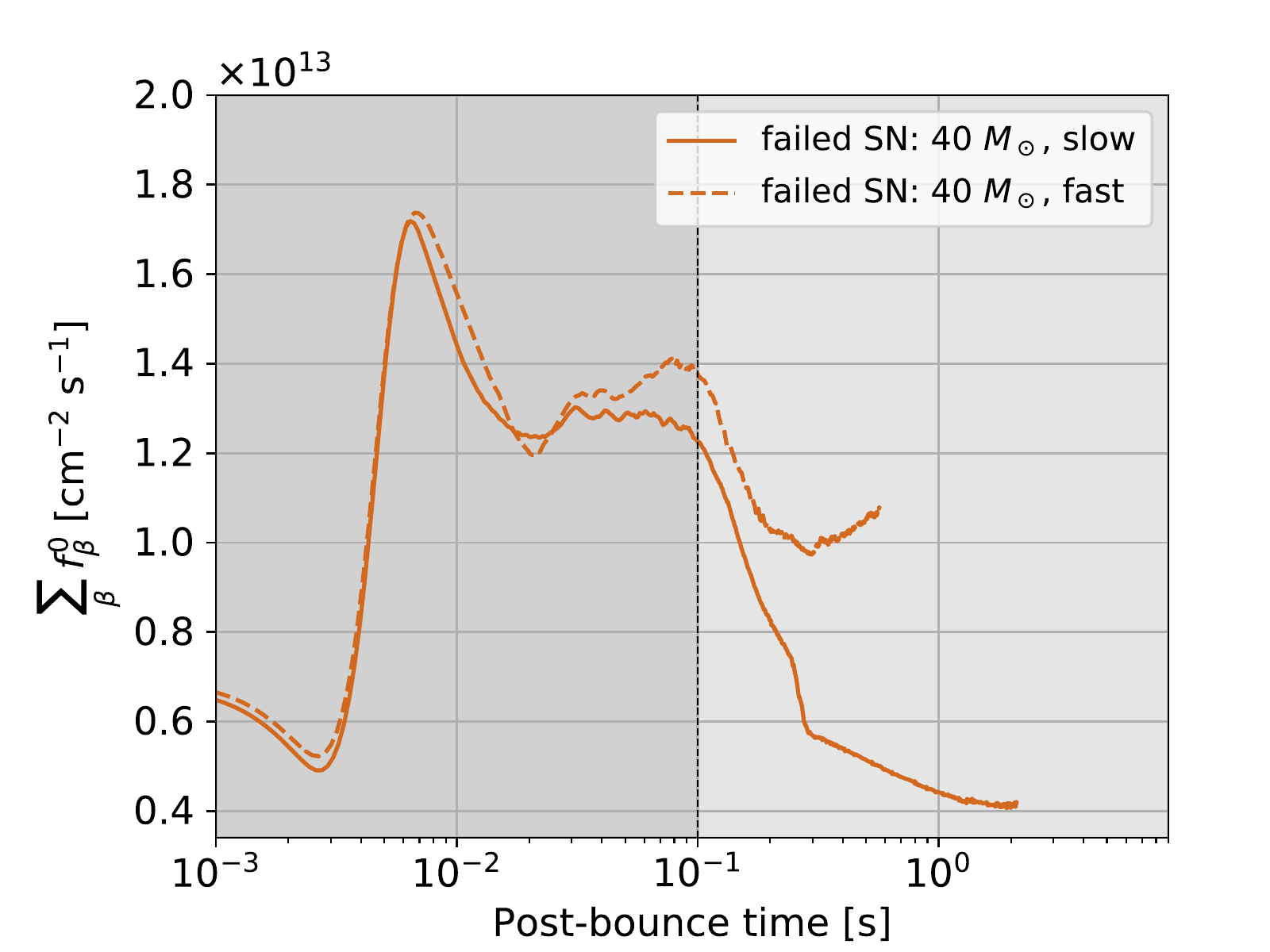}
\caption{Temporal evolution of the total neutrino flux (defined as in Eq.~\ref{eq:nuflux} for each $\nu_\beta$ and summed over all six neutrino flavors) for a stellar collapse occurring at $10$~kpc. {\it Left panel:} Neutrino fluxes for our benchmark $9.6$ and $27\ M_\odot$ core-collapse SN models  with LS220 and SFHo EoSs. The three shaded regions highlight the neutronization burst, the accretion phase, and the Kelvin-Helmholtz cooling phase, from left to right respectively. {\it Right panel:} Neutrino fluxes for our  $40\ M_\odot$ black hole forming collapses with slow and fast accretion (slow and fast failed SN, respectively). The black hole formation suddenly halts the neutrino signal.}
\label{fig:models} 
\end{figure*}
The resultant temporal evolution of the neutrino flux introduced in Eq.~\ref{eq:nuflux} summed over all six flavors, for $d=10$~kpc, and for the six benchmark SN models is shown in Fig.~\ref{fig:models} for the standard core-collapse SN models (CC-SN) on the left and for the black hole stellar collapses (failed SN) on the right.  As visible from the left panel of Fig.~\ref{fig:models}, the neutrino signal lasts for $\mathcal{O}(10)$~s, but it drops considerably after a few seconds. As highlighted by the gray-shaded regions, the overall neutrino signal can be divided in three time windows: the neutronization burst, the accretion phase, and the Kelvin-Helmholtz cooling phase. The neutronization burst originates as the shock wave is moving outwards through the iron core and free protons and neutrons are produced due to iron dissociation.  The rapid electron capture by nuclei and free protons is responsible for generating a $\nu_e$ burst lasting for about $\sim 0.05$~s. In this phase, the neutrino signal is mildly dependent on the SN mass and nuclear EoS [$\mathcal{O}(1-10\%)$]~\cite{OConnor:2018sti}.

The neutronization burst is followed by the accretion phase lasting until $\sim 1$~s. The shock after losing all its energy by dissociating iron nuclei stalls, and neutrinos are thought to provide fresh energy to the shock to finally revive it and trigger the explosion according to the delayed neutrino heating mechanism. This eventually leads to the explosion after few hundreds of milliseconds. 

After, the Kelvin-Helmholtz phase occurs. As the newly born proto-neutron star cools and deleptonizes, the neutrino flux gradually decreases  and the neutrino emission properties tend to become similar across all flavors. 
Section~2.4.6 of Ref.~\cite{Mirizzi:2015eza} provides a detailed description of the dependence of the neutrino emission properties from the EoS. However, one can notice that the dependence of the neutrino fluxes from the EoS is weaker than their dependence on the progenitor mass.

The right panel of Fig.~\ref{fig:models} shows the neutrino flux expected from failed SNe for comparison. The first clear difference between the two SN subsets is the duration of the neutrino signal. Black hole formation halts the neutrino emission after $0.57$~s for the model s40s7b2c (fast failed SN) and after $2.1$~s for the model s40c  (slow failed SN). Moreover, the latter model displays lower mass accretion rate (see Ref.~\cite{Mirizzi:2015eza} for more details).  

It is worth noticing that 1D SN models describe the overall neutrino emission properties well, but do not carry signatures of the hydrodynamical instabilities that may affect the neutrino signal during the accretion rate, see e.g.~Refs.~\cite{Tamborra:2014hga,Walk:2019miz,Janka:2016fox}. However, in this work, we are interested in reconstructing the general properties of the SN neutrino burst, and neglect any small-scale feature affecting the neutrino signal.

The total flux summed over six flavors in Fig.~\ref{fig:models} is insensitive to flavor conversions. However, neutrinos change their flavor while they propagate through the stellar envelope as well as on their way to Earth. In the proximity of the SN core, the neutrino density is so high that neutrino-neutrino interactions are believed to be dominant giving rise to non-linear effects in the flavor evolution history. 
At larger radii, neutrinos undergo Mikheev-Smirnov-Wolfenstein (MSW) resonant conversions because of interactions with the matter background. Turbulences or large stochastic fluctuations of the matter density in the SN envelope can affect the neutrino flavor distribution. We refer the reader to Refs.~\cite{Mirizzi:2015eza,Horiuchi:2017sku,Chakraborty:2016yeg} for recent reviews on the topic and references therein. 

The total neutrino flux summed over all six flavors may still be affected by non-standard physics. This is the case of  heavy and light sterile neutrinos~\cite{Suliga:2019bsq,Wu:2013gxa,Mastrototaro:2019vug}, secret neutrino  interactions~\cite{Kolb:1987qy,Shalgar:2019rqe}, neutrino decay~\cite{deGouvea:2019goq,Kolb:1988pe}, beyond the Standard Model light particles~\cite{Sung:2019xie}, and non-standard neutrino interactions~\cite{Stapleford:2016jgz,EstebanPretel:2007yu}. Exploiting CE$\nu$NS as another detection channel for SN neutrinos may open a new portal for probing the Standard Model.

\section{Neutrino detection in a lead-based cryogenic detector}\label{sec:CEnuNS}

In this Section, we focus on the intrinsic potential of exploiting CE$\nu$NS in a Pb-based cryogenic detector and introduce the main features of our proposed neutrino telescope RES-NOVA. A brief overview of the expected backgrounds and the detector energy response is also provided.

\subsection{The RES-NOVA detector concept}
RES-NOVA\footnote{Latin word for {\it new thing}. This name is meant to identify a {\it new class} of Neutrino Observatories.} is a proposed experiment for the detection of neutrinos from astrophysical sources, based on the employment of archaeological Pb as sensitive detector component.
This valuable material is already available  in Italy~\cite{Fiorini,PersComm}, at the  Gran Sasso National Laboratories  (about 4~ton) and  at the Archaeology Department of the Ministry for Cultural Assets of Cagliari (about 33~tons). 
Securing archaeological Pb allows for a prompt project realization. 

The RES-NOVA concept relies on  state-of-the-art technology of cryogenic detectors, which has already proved its potential. No specific R\&D would be needed for the detector realization, other than a small scale demonstrator. Hence, we expect a commissioning time of about three years, during which funding will be secured for the detector implementation and study. 

\subsection{Coherent elastic neutrino-nucleus scattering in a lead-based cryogenic detector}

CE$\nu$NS is a NC process, equally sensitive to all neutrino flavors. This feature makes CE$\nu$NS a complementary approach to other conventional techniques for neutrino detection which are instead mostly sensitive to $\bar\nu_e$ or $\nu_e$.  In addition, as already pointed out in \cite{Drukier:1983gj}, CE$\nu$NS offers a wealth of applications in neutrino physics being a threshold-less process, thus sensitive to the full SN neutrino signal~\cite{Lang:2016zhv,Horowitz:2003cz}. 

The interaction cross-section of CE$\nu$NS can be easily computed by Standard Model basic principles for all neutrino energies and it has been  measured  experimentally recently~\cite{Akimov:2017ade}. The cross-section is~\cite{Freedman:1977xn}:
\begin{eqnarray}
\label{eq:xsec}
\frac{d\sigma}{d E_R} &=& \frac{G^2_F m_N}{8 \pi (\hslash c )^4} \left[(4\sin^2 \theta_W -1 ) Z + N\right]^2   \nonumber\\ 
&\; & \left(2- \frac{E_R m_N}{E^2}\right) \cdot |F(q)|^2\ ,
\end{eqnarray}
where $G_F$ is the Fermi coupling constant, $\theta_W$ the Weinberg angle, $Z$ and $N$ the atomic and neutron numbers of the target nucleus, while $m_N$ its mass, $E$ the  energy of the incoming neutrino and $E_R$ the recoil energy of the nucleus. The last term of the equation, $F(q)$, is the elastic nuclear form factor at momentum transfer $q=\sqrt{2E_R m_N}$. It represents the distribution of the weak charge within the nucleus and for small momentum transfers its value is close to unity. The parameterization of $F(q)$ follows the model of Helm~\cite{Helm:1956zz}; for an exact evaluation of $F(q)$ see~\cite{PhysRevC.85.032501}.

Elastic neutrino scattering on Pb nuclei is coherent, if the energy of the incoming particles is $E \lesssim 30$~MeV. For this reason, Pb can be considered an efficient  target for SN neutrinos, although SN neutrinos may have higher energies ($[1,50]$~MeV, see Sec.~\ref{sec:nusignal}). For the sake of accuracy, we also take into account possible non fully-coherent interactions through the evaluation of $F(q)$ for different momentum transfers.
It is worth mentioning that having a target nucleus with  high $N$ increases the cross-section, and if the interaction is coherent we have a further enhancement: $\sigma_{\rm{CE\nu NS}}\propto N^2$.

\begin{figure}
\centering
\includegraphics[width=0.48\textwidth]{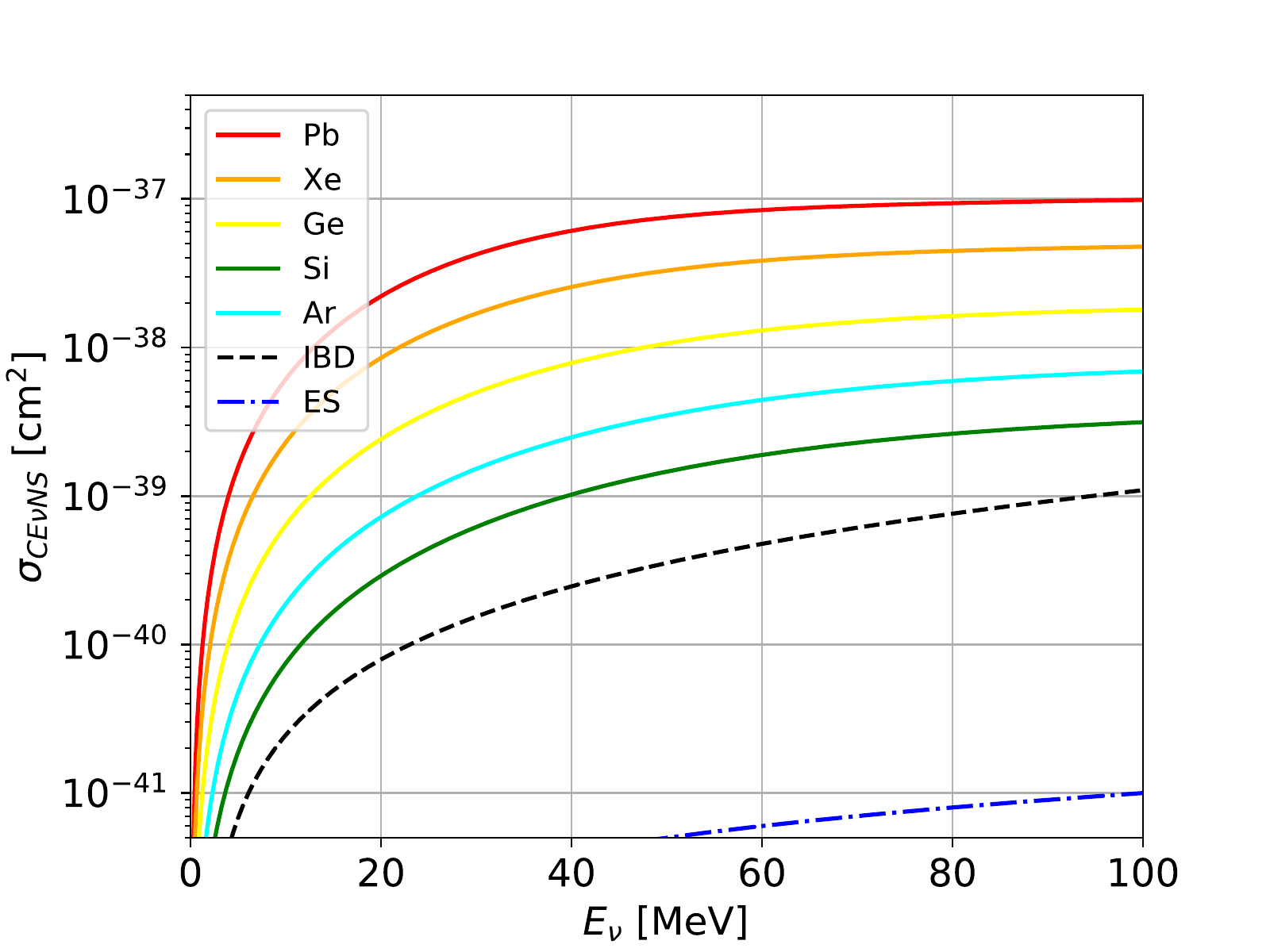}
\caption{Coherent elastic neutrino-nucleus scattering (CE$\nu$NS) cross sections as a function of the energy of the incoming neutrino for different target nuclei. The dashed lines show the inverse-beta decay (IBD) and neutrino elastic scattering on electrons (ES) cross-sections for comparison. Given the high cross-section, CE$\nu$NS has the potential to provide large statistics with small detector volumes.}
\label{fig:xsec}
\end{figure}

Figure~\ref{fig:xsec} displays the CE$\nu$NS cross-section for different nuclei frequently used for the realization of neutrino detectors. From an experimental point of view, Pb is the best target candidate to be used for neutrino detection via CE$\nu$NS thanks to the high $N$, thus high cross-section, and its nuclear stability, intrinsic low backgrounds~\cite{Beeman:2012wz}. Furthermore, CE$\nu$NS has an interaction cross-section much higher than the conventional IBD and ES channels. CE$\nu$NS is about four orders of magnitude greater than other NC processes, namely ES. This is a great advantage for detectors exploiting CE$\nu$NS, in fact they can potentially  achieve higher statistics with smaller detector volumes.

The  signature produced by astrophysical neutrinos interacting on Pb via CE$\nu$NS is the nuclear recoil of few keV of energy. In order to evaluate the time and energy distribution of the neutrino induced events we compute the total number of expected events in the detector ($N_{\rm{exp}}$). This is done by integrating the product of the total neutrino flux emitted over $10$~s (Eq.~\ref{eq:nuflux}) and the CE$\nu$NS differential interaction cross-section (Eq.~\ref{eq:xsec}) over the relevant incident neutrino energy range, and by multiplying the resulting function for the total number of target nuclei ($N_{\rm{Pb}}$): 
\begin{eqnarray}
N_{\rm{exp}} &=&  \int dt\int_{E_{R\rm{min}}} \frac{d^2N}{dE_R\; dt} dE_R =\\ \nonumber
&=&\sum_{\alpha} N_{\rm{Pb}} \int dt \int_{E_{\rm{min}}} f_\alpha^0(E,t) \; \frac{d\sigma}{dE_R}  \;dE\ , \label{eq:rate}
\end{eqnarray}
where the sum is over all six neutrino flavors.

The best available experimental technique for the detection of keV nuclear recoils in Pb-based detectors is the cryogenic one~\cite{Beeman:2012wz}. Low temperature cryogenic detectors will bring, for the first time in a neutrino telescope, the following unique features:
\begin{itemize}
\item[-] fully active volume. The entire detector is sensitive to particle interactions, and no volume fiducialization is needed, unlike all other adopted technologies (e.g., water Cherenkov detectors, noble-liquid TPCs);
\item[-] per-mille energy resolution over a wide energy range~\cite{Abdelhameed:2019hmk,Alfonso:2015wka,Alessandria:2012ha}, from few eV up to some MeV, allowing the identification of any structure in SN neutrino recoil spectra;
\item[-] scalability to large detector volumes~\cite{Alduino:2017ehq};
\item[-] operation of different detector compounds~\cite{Beeman:2012wz,Casali:2013zzr,Beeman:2011kv,Pattavina:2018nhk}, without being limited by the technology to a single compound (e.g. water, liquid-Ar/Xe).
\item[-] active background suppression techniques, such as scintillating compounds for particle identification (e.g. e$^-$/$\gamma$, $\alpha$, neutrons)~\cite{Azzolini:2019tta,Beeman:2012wz}.
\end{itemize}
For the realization of the RES-NOVA detector different crystal absorbers are taken into consideration: pure Pb crystals or Pb containing crystals, as PbWO$_4$ or PbMoO$_4$. These crystals were already operated as cryogenic calorimeters with excellent performance~\cite{Beeman:2012wz,Pattavina:2019pxw,Pattavina:2020ota}.\\

In order to show the potential of the proposed Pb-based cryogenic detector technology, in the following we will consider the $27\ M_\odot$ model with LS220 EoS as our {\it reference SN model}. This will be used for the estimation of the optimal detector parameters, namely energy threshold and detector volume (mass). We will focus only on the initial $10$~s of the neutrino signal after the core bounce.

\begin{figure}[t!]
\centering
\includegraphics[width=0.45\textwidth]{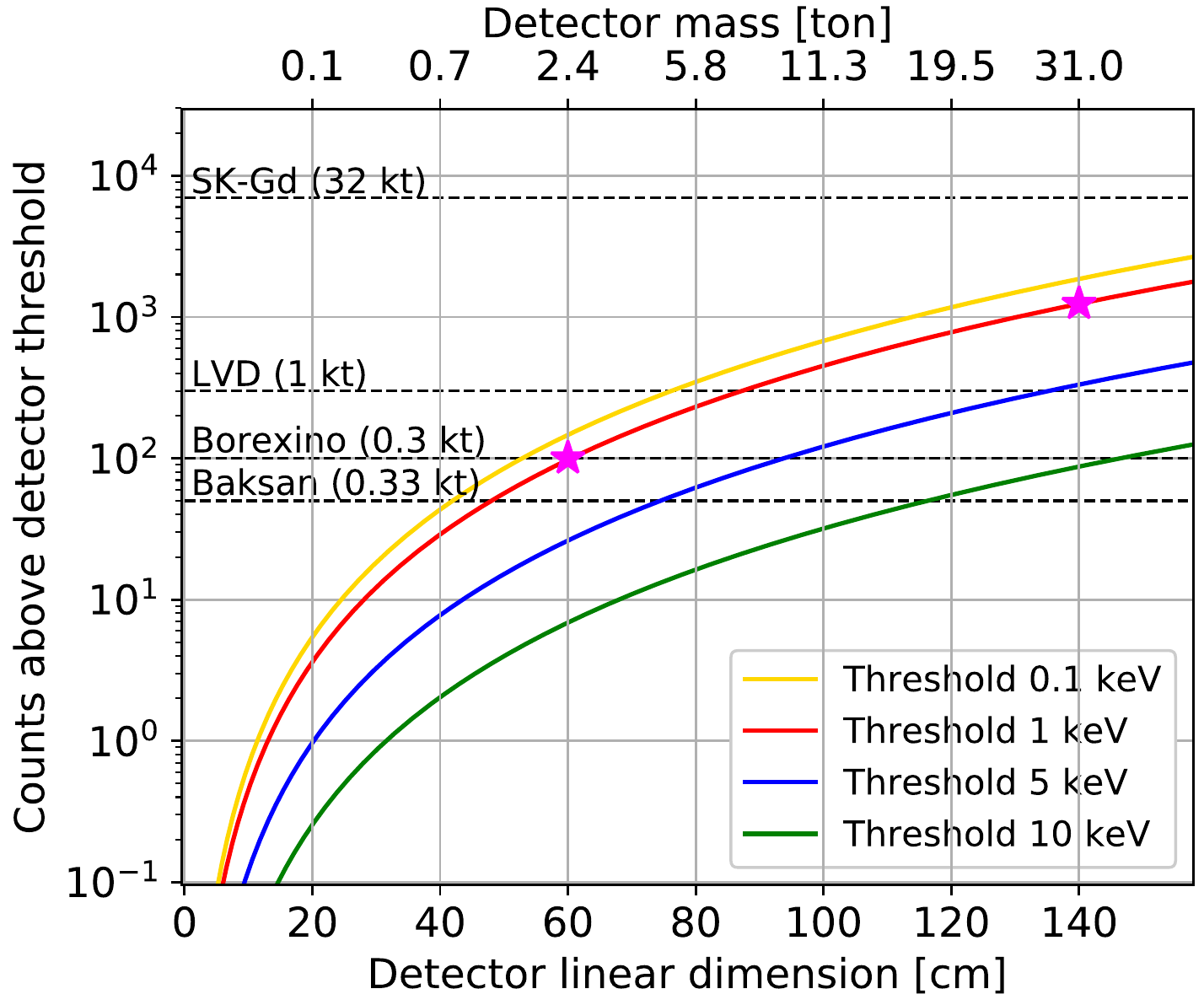}
\caption{Number of events for different linear dimensions (or effective mass) of the detector and different energy thresholds for our benchmark $27\ M_\odot$ SN model with LS220 EoS integrated over $10$~s after the core bounce. The horizontal lines represent the signal observed in dedicated  neutrino observatories for the same SN input~\cite{Scholberg:2012id}; in parenthesis the respective detector masses are shown. A Pb-based detector with linear dimensions of tens of cm can achieve a number of detected events similar to $\mathcal{O}(\mathrm{kton})$ flavor-dependent detectors. The two markers at $60$~cm and $140$~cm represent the sensitivity of the two first phases of the proposed RES-NOVA project for our fiducial energy threshold.}
\label{fig:size} 
\end{figure}

Figure~\ref{fig:size} shows the integrated total number of detected events ($\int d^2N/(dE_R dt) dE_R dt$) for different detector energy thresholds as a function of the detector linear dimension. We define the detector linear dimension as the length of the side of a cube whose volume is equivalent to the total detector active volume (Pb-based crystals). A detector with linear dimensions of tens of cm can achieve similar sensitivity (comparable number of detected events) to $\mathcal{O}(\mathrm{kton})$ detectors. With the additional advantage that the proposed Pb detector is equally sensitive to all neutrino flavors.
As shown in the secondary (top) x-axis of Fig.~\ref{fig:size}, the high density of Pb can potentially allow to achieve large detector masses with small detector volumes in comparison with dedicated neutrino detectors. Thus, the proposed technology has a great potential for detector upgrades to larger volumes.
 
The RES-NOVA research program will be divided in three main phases (see Tab.~\ref{tab:detector}), with increasing detector volumes and sensitivities. In Fig.~\ref{fig:size}, the two initial phases of the program are highlighted by two markers. For this estimation we took into account our benchmark core-collapse SN at 10~kpc. In the first phase, the RES-NOVA detector will be able to detect about 100 SN neutrino events with a total detector active volume of ($60$~cm)$^3$ and an energy threshold of only $1$~keV. For the sake of comparison and demonstration of the effective detector miniaturization, the large active volume of Borexino (sphere with a diameter of 9~m) can detect, from a similar SN, about 30~NC neutrino interactions~\cite{Cadonati:2000kq}. The second phase foresees a detector scaling to ($140$~cm)$^3$ of the total volume without any other technological advancement;  the third phase would consists in installing multiple ($140$~cm)$^3$ detectors in $15$ different sites. Table~\ref{tab:detector} summarizes the main detector characteristics of the RES-NOVA research program.
 
\begin{table}[b]
\centering
\caption{Detector characteristics for the three phases of the RES-NOVA research program, see main text for more details. The background index column refers to the background in the $1$--$40$~keV region of interest.}
\begin{tabular}{|c|c|c|c|c|}
\hline
 &
  \begin{tabular}[c]{@{}c@{}}\Tstrut Linear\\ dimension\Bstrut\end{tabular} &
  \begin{tabular}[c]{@{}c@{}}Detector\\ mass\end{tabular} &
  \begin{tabular}[c]{@{}c@{}}Energy\\ threshold\end{tabular} &
  \begin{tabular}[c]{@{}c@{}}Background\\ index\end{tabular} \\ \hline
\Tstrut RN-1 & 60 cm\Bstrut & 2.4 t & 1 keV & 0.1 c/keV/t/10 s \\ \hline
\Tstrut RN-2 & 140 cm\Bstrut & 31 t  & 1 keV & 0.1 c/keV/t/10 s \\ \hline
\Tstrut RN-3 & 15$\times$ 140 cm\Bstrut & 465 t & 1 keV & 0.1 c/keV/t/10 s \\ \hline
\end{tabular}
\label{tab:detector}
\end{table}

In summary, running a Pb based cryogenic detector employing CE$\nu$NS will be of great importance given the high CE$\nu$NS interaction cross-section and advantages of the cryogenic technique. This innovative approach has the potential to allow an important downsizing of the overall experimental volume, enabling future upgrades of the set-up for higher sensitivities. As we will discuss in the next Section, this is possible only if high-purity archaeological Pb is employed and if the Pb detectors are operated as low temperature calorimeters.

\subsection{RES-NOVA detector design}\label{sec:design}

The RES-NOVA detector has a modular design. This ensures a fast and effective detector upgrade, without the need of specific technological improvements.
\begin{figure}[b]
\centering
\includegraphics[width=0.45\textwidth]{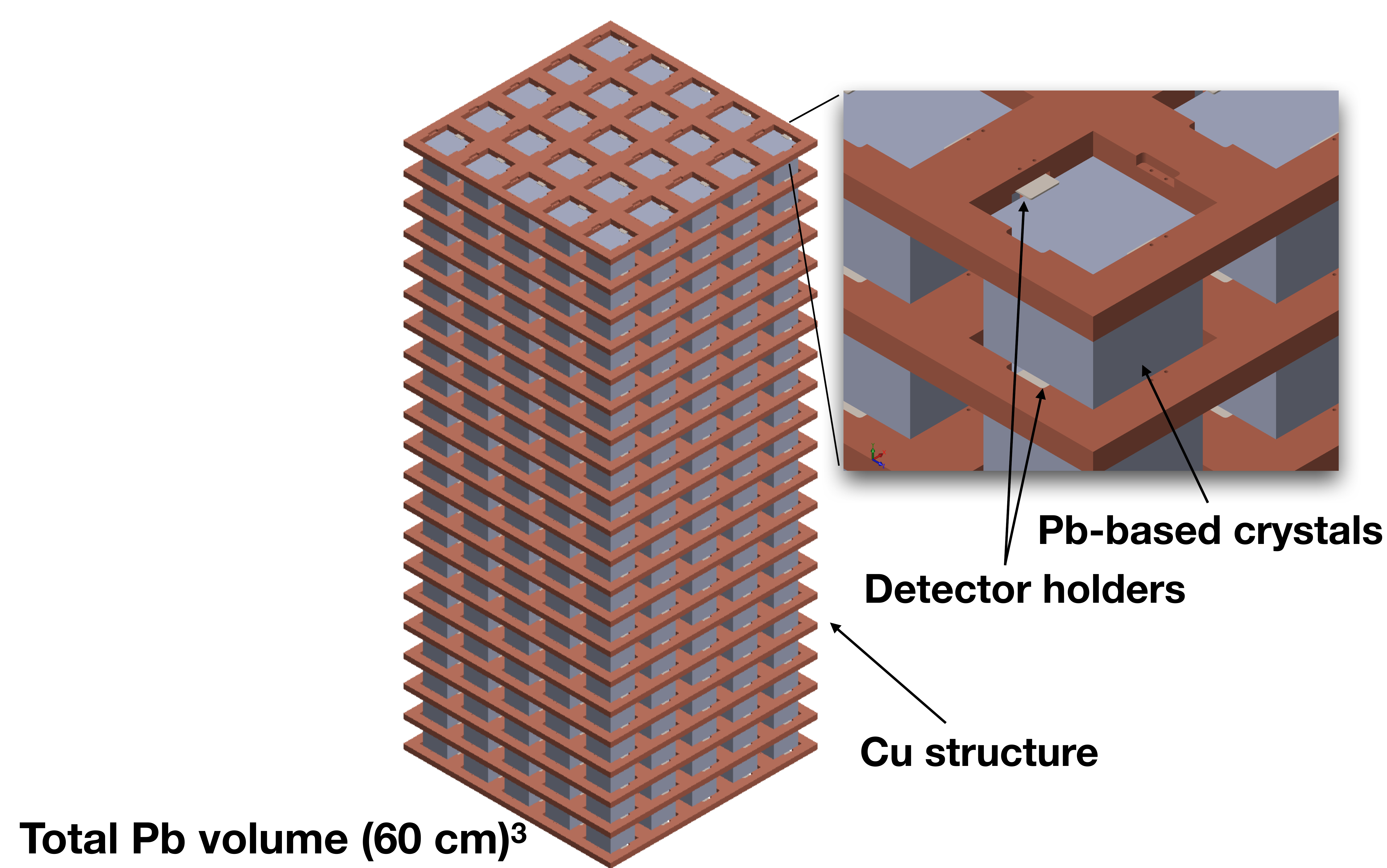}
\caption{Schematic drawing of the Phase-1 detector, total volume ($60$~cm)$^3$. The detector is composed by $20$ floors of 25 crystals each, arranged in a tightly packed configuration. Copper is used, at the same time, as detector holding structure and thermal bath for the detector cool-down. The inset shows a zoom-in of the detector single module.}
\label{fig:sketch} 
\end{figure} 
In the first phase (RES-NOVA Phase-1, RN$^1$), the detector has a volume small enough to fit in the commercially available cryogenic infrastructures, and it consists of 500 detectors with total linear dimension of the array of $60$~cm; note that the linear dimension refers to the effective active detector volume. A preliminary design of the RN$^1$ experimental apparatus is sketched in Fig.~\ref{fig:sketch}. The second phase (RES-NOVA Phase-2, RN$^2$) is expected to replicate the  RN$^1$ design for  $\sim$12 times and should fit inside a larger experimental facility, similar to the one currently operated by CUORE~\cite{Alduino:2019xia}, the largest cryogenic infrastructure for low temperature detectors. The infrastructure will host 6000 crystals with total linear dimension of $140$~cm. Finally,  the ultimate phase of this  research program (RES-NOVA Phase-3, RN$^3$) foresees  $15$ detectors, with technical features similar to the ones of RN$^2$,  installed in deep-underground laboratories world-wide, creating a Pb-based network of SN neutrino observatories.

We foresee to operate detectors with an energy threshold of $1$~keV, which is within the reach with the available technologies. The current lowest energy thresholds achieved with low temperature detectors are at the level of tens of eV, as demonstrated by the CRESST~\cite{Abdelhameed:2019hmk}, the EDELWEISS~\cite{Armengaud:2019kfj} and CDMS~\cite{Agnese:2017jvy} experiments. These experiments are running two classes of thermal sensors: Transition Edge Sensors (TES) and Neutron Transmutation Doped Ge thermistors (Ge-NTD). These are two very different read-out technologies for the measurement of temperature variations induced by particle interactions~\cite{Pirro:2017ecr}. Nevertheless, both  can  be implemented in the proposed RES-NOVA project.
Large arrays of cryogenic detectors were already shown to be feasible, as demonstrated by the CUPID-0~\cite{Azzolini:2018tum} and the CUORE~\cite{Alduino:2017ehq} experiments, running respectively 57 and 988 calorimeters.

The technology proposed for the realization of RES-NOVA is already in place. Nevertheless, the effective feasibility of merging all the necessary technologies in one single experiment needs to be demonstrated.
In order to show the potential of the proposed RES-NOVA research program, in the following, we will forecast the expected background level and SN neutrino event rate in a lead-based detector.

\subsection{Expected backgrounds}

In the field of cryogenic detectors for rare event investigations~\cite{Pirro:2017ecr}, the main limitation to the experimental sensitivity is the background produced by cosmic-rays induced particles (e.g. neutrons) and the detector intrinsic radiopurity, namely the concentration of $^{235/238}$U and $^{232}$Th decay chain products in the absorber. While the first background source can easily be  mitigated or suppressed by installing the experimental set-up in deep underground laboratories and equipping them with highly efficient vetoes, the second one needs a different approach.
The materials that should be used for the detector realization should  be selected for their intrinsic radiopurity. For this reason, a thorough screening of the concentration of $^{235/238}$U and $^{232}$Th decay chain radio-nuclides is mandatory before the detector realization.

As already mentioned, Pb is the ideal element to be used as detector target for the detection of $\mathcal{O}(10~\mathrm{MeV}$) neutrinos. Unfortunately, the main background source in Pb containing detectors is $^{210}$Pb. This is a natural radioactive isotope produced by the $^{238}$U decay chain. It has a half-life of $22.3$~yr and it $\beta^-$ decays  with a low energy Q-value of $63$~keV, exactly around our region of interest (RoI). For these reasons, $^{210}$Pb is considered to be the most dangerous isotope when investigating low-energy rare events  (e.g.~dark matter searches and CE$\nu$NS related searches). Whenever $^{210}$Pb contaminates a detector (whether it is a cryogenic one or not), it can only be partially removed by means of advanced cleaning and purification techniques~\cite{Alessandria:2012zp,Giampa:2018pbh}.

Here, we propose to overcome all these issues by running a detector made of archaeological Pb. If the age of Pb is old enough (e.g.~Roman Pb is more than $2000$~yr old) the concentration of $^{210}$Pb, and its decay products, is strongly suppressed. The outstanding radiopurity of an archaeological Roman Pb cryogenic detector has been recently demonstrated in Ref.~\cite{Pattavina:2019pxw}. There, the lowest limit on the concentration of $^{210}$Pb in archaeological Pb samples was measured:  $< 2.5\times 10^{-18}$~g/g ($<715$~$\mu$Bq/kg), 4-orders of magnitude better than any other low-radioactivity Pb sample~\cite{Pattavina:2019pxw}.
The concentration of other radio-nuclides in archaeological Pb, namely $^{235/238}$U and $^{232}$Th, was also investigated in other works; also the concentration of these radio-nuclides were found to be extremely low, $< 4\times 10^{-12}$~g/g and $< 11\times 10^{-12}$~g/g~\cite{Alduino:2017qet}, respectively.
Our novel approach consists of using this valuable material not as detector passive shielding but as active detector component.

\begin{figure}
\centering
\includegraphics[width=0.48\textwidth]{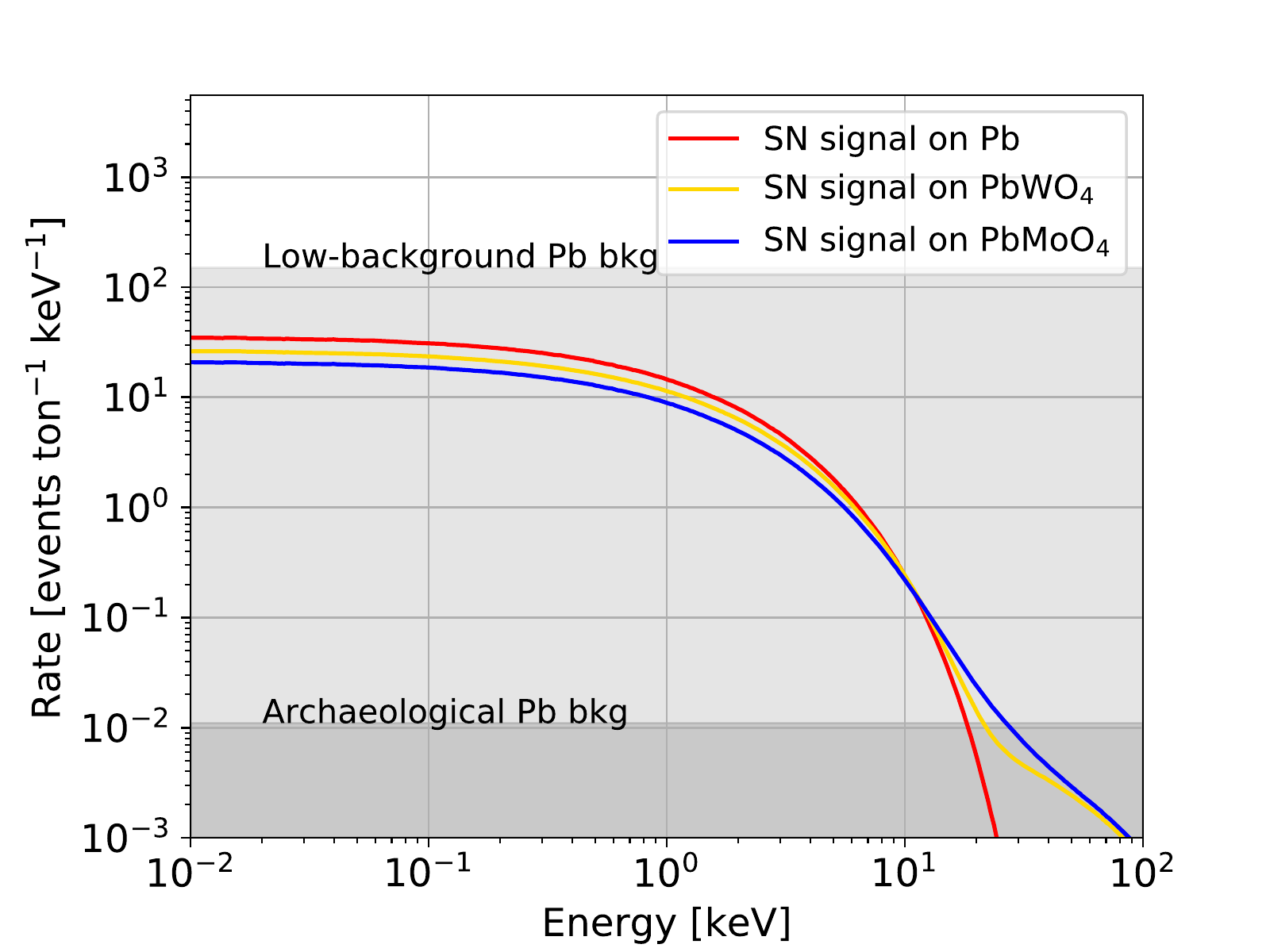}
\caption{Expected total number of SN neutrino events  as a function of the recoil energy (keV$_{\rm{ee}}$ $\simeq$ keV$_{\rm{nr}}$) for a SN burst at $10$~kpc in a detector made of Pb (in red), PbWO$_4$ (in yellow) and PbMoO$_4$ (in blue). The all-flavor and time-integrated neutrino signal for the $27\ M_\odot$ SN model with LS220 EoS has been adopted as input. The light and dark grey areas represent the expected background levels induced by low-background Pb~\cite{Heusser} and by archaeological Pb~\cite{Pattavina:2019pxw}, respectively. The background bands refer to electron and nuclear recoil interactions from $^{210}$Pb, $^{238}$U and $^{232}$Th decay chains.}
\label{fig:spectrum} 
\end{figure}

The sensitivity of an archaeological Pb-based cryogenic detector to SN neutrinos is shown in Fig.~\ref{fig:spectrum}. The detector energy response to the SN {\it reference signal} ($27\ M_\odot$ model with LS220 EoS at 10~kpc) is shown. As it will be discussed in the following sections, the energy scale is in keV, given that there is no detector quenching (keV$_{\rm{ee}} \simeq$ keV$_{\rm{nr}}$). The detector counting rate, normalized per detector unit mass and for a neutrino signal duration of $10$~s, is plotted as a function of the recoil energy measured by the detector. In the same plot, the expected background level from a conventional low-background Pb~\cite{Heusser} and from an archaeological Pb~\cite{Pattavina:2019pxw} sample are also shown. These represent the total background from electron recoils (from $\beta$/$\gamma$ interactions) and nuclear recoils (from $\alpha$ decays). Preliminary Monte Carlo simulations confirm that the background can be assumed flat over the RoI. Notably, our estimation of the archaeological Pb background is conservative and based on the current limits on its impurity concentrations. Possibly, this contribution will be even lower. A statistically significant detection of SN neutrinos is possible only if the Pb background is minimized. In this respect, archaeological Pb ensures an outstanding signal-to-noise ratio over the entire RoI.

The potential of this new approach (ultra-low background, fully-active detector and high interaction cross-section) is demonstrated by the detector miniaturization. In fact, even for a detector with linear dimensions of about $40$~cm (equivalent to $1$~ton of mass), the SN neutrino signal would be about 3-orders of magnitude higher than the expected background. 

We  stress that a detailed background model is mandatory for a precise computation of the RES-NOVA sensitivity to different neutrino signals. This should also include the contribution from external neutron sources (e.g., muon-induced neutrons), and from detector ancillary components (e.g., radioactive surface contamination). The latter is known to be one of the most relevant background sources in cryogenic detectors operated in Astroparticle Physics~\cite{Alduino:2017qet,Kuzniak:2012zm}.
However, in this paper, we explore the detection potential and consider only the background
induced by the Pb itself,  because  this is as an irreducible background. Other background sources can be mitigated and reduced by means of a proper detector design and shielding; nevertheless, a full-comprehensive detector design is needed  in order to take into account these aspects and will be the main focus of future papers.

\subsection{Detector energy response}

The detector energy response to a time-integrated SN signal and the expected background are shown in Fig.~\ref{fig:spectrum}. 
We assumed as detector energy resolution of 0.2~keV\footnote{We conservatively assume the detector energy resolution to be 1/5 of the detector energy threshold: $E_{\rm{thr}}=5\;\sigma$. Thus, if $E_{\rm{thr}}=1$~keV,  $\sigma=0.2$~keV} over the entire RoI.

The advanced detector performance is achievable thanks to the extremely low energy of the information carriers in solid state cryogenic detectors, which are phonons with energy of few $\mu$eV. The energy distribution of the phonons depends on the operating temperature of the detector, which in our case is $\mathcal{O}(\mathrm{mK})$. The energy deposited in the absorber by one neutrino interaction is of the order of few keV. So for a single neutrino interaction about $10^9$ phonons are generated in the detector. The small statistical and thermodynamic fluctuations of the system allow to achieve an energy resolution at the permil scale over a broad energy range. This experimental approach makes possible the operation of almost ideal calorimeters able to measure the entire energy deposited in the detector irrespectively of the type of interacting particle and with very limited uncertainties related to the energy reconstruction (e.g. nuclear recoil quenching).

The detector response for the three compounds considered (pure Pb, PbWO$_4$ and PbMoO$_4$) is shown in Fig.~\ref{fig:spectrum}. There is no relevant difference in the detector counting rates for the different compounds, except for the higher energy component. Compounds containing oxygen feature a higher energy tail of the SN signal, due to the lightness of the element, hence higher transferred momenta.

In Fig.~\ref{fig:spectrum}, only the rate from the CE$\nu$NS channel is shown. Other channels may give a contribution to the detector counting rate, such as the CC interactions on Pb but in a different region of interest [$\mathcal{O}$(MeV)]. The latter is the main detection channel exploited in the HALO experiment~\cite{Duba_2008}. However, the expected counting rate from this channel is more than two orders of magnitude lower than the CE$\nu$NS one. For a similar {\it reference SN model}, the yield would be about 0.2~events/ton~\cite{Vaananen:2011bf}. 

The advantage of considering  Pb-based compounds consists of  facilitating the  establishment of  scintillating cryogenic detectors~\cite{Beeman:2012wz,Pattavina:2020ota}. This experimental technique~\cite{Pirro:2017ecr} can enable a particle identification and discrimination (nuclear recoils vs.~$\beta$/$\gamma$ interactions), leading to a powerful background suppression. This point is particularly interesting for the study of the  diffuse SN neutrino background, as discussed in Sec.~\ref{sec:dsnb}. For the sake of simplicity, in the following we will only consider a detector made of pure Pb crystals.

\section{Reconstruction of the neutrino emission properties for a galactic burst}\label{sec:SNburst}
In this Section, we explore the possibility of detecting a SN in our Galaxy through a lead-based cryogenic detector. As representative case of study, we focus on the reconstruction of the neutrino signal as a function of time and discuss the potential of reconstructing the neutrino emission properties.  At last, we discuss the SN neutrino detection significance of RES-NOVA  for extragalactic bursts.

\subsection{Reconstruction of the supernova neutrino light curve}

\begin{figure}
\centering
\includegraphics[width=0.48\textwidth]{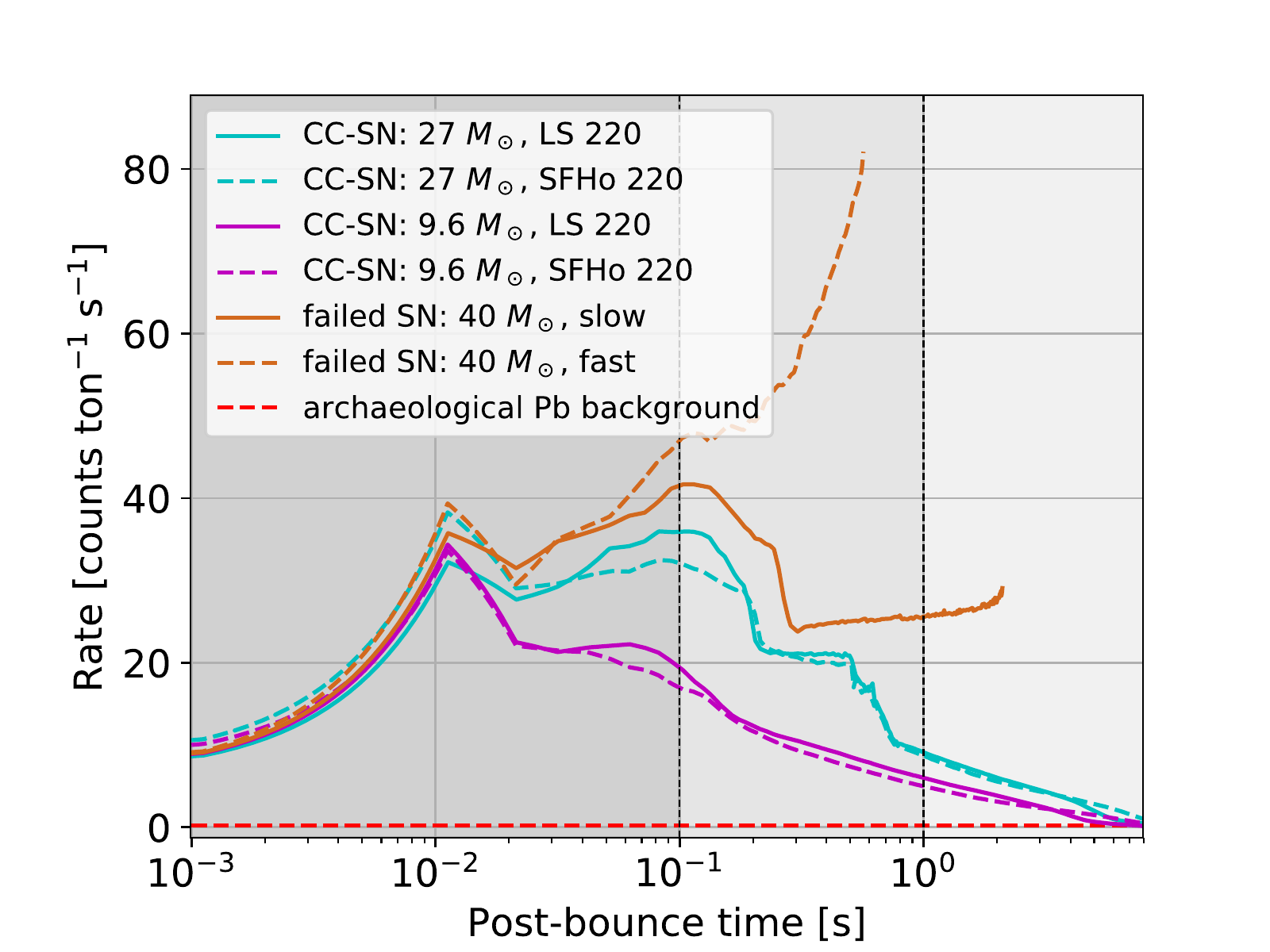}
\caption{Recoil rate as a function of the post-bounce time per unit of detector mass for six SN models. Threshold effects are not considered. The SN is assumed to be at a distance of $10$~kpc. The time profile of the event rate is convolved with the detector time response. The gray shadowed regions mark the three different phases of the SN neutrino signal (neutronization burst, accretion phase, and cooling phase). The event rates of  different SN models are easily distinguishable from each other, although there is a smaller variation of the expected rate due to the nuclear EoS.}
\label{fig:recoils} 
\end{figure}

A Galactic SN will be a once-in-a-lifetime opportunity. Hence, we should be able to extract as much as possible information from the burst. We compute the expected counting rate of the RES-NOVA detector as a function of time by integrating Eq.~\ref{eq:rate} with respect to $E_R$ over the RoI, from the detector energy threshold to $40$~keV. Furthermore, the computed $dN/dt$ is smeared according to the detector time resolution, which is conservatively assumed to be $100$~$\mu$s. The detector response to the different SN models in Fig.~\ref{fig:models} is shown in Fig.~\ref{fig:recoils}, where the SN burst is  assumed to occur at a distance of $10$~kpc from Earth. The detector counting rate is normalized per detector mass unit, thus it is sufficient to multiply the rate for each RES-NOVA detector mass to compute the expected total number of SN neutrino events as a function of time.

The main spectral features of Fig.~\ref{fig:models}  also appear in Fig.~\ref{fig:recoils}. All the models show the neutronization peak at $\sim 10$~ms, followed by a decrease of the detector counting rate. The only exception is the fast failed SN model characterized by an increasing counting rate. The RES-NOVA technology, even in its first stage, is potentially able to clearly identify the type of SN event (i.e., core-collapse or failed burst) by looking at the time distribution of the detector counting rate, namely the last bin which shows a neutrino event.

In order to precisely estimate the SN mass and EoS, a high statistics of neutrino events is crucial. For this reason, the larger is the RES-NOVA detector mass the more information will be retrieved. This is the focus of the next section.

\subsection{Supernova discrimination}

We now explore the potential of a cryogenic Pb-detector to investigate the properties of a SN event by relying on the time evolution of the related neutrino light curve (see Fig.~\ref{fig:recoils}). We determine the total number of expected events by integrating $dR/dt$ over the relevant 
post-bounce time interval:
\begin{equation}
    N_{\mathrm{exp}} = \int \frac{d^2N}{dE_R\; dt}  \ dt \ dE_R = \int \frac{dN}{dt}\ dt\ ,
    \label{eq:Nexp}
\end{equation}
and the probability density function for an event to occur at the time ${t}$:
\begin{equation}
    p(t) = \frac{1}{N_{\mathrm{exp}}} \frac{dN}{dt}\ .
    \label{eq:Pt}
\end{equation}
Thanks to Eqs.~\ref{eq:Nexp} and \ref{eq:Pt}, we  generate Monte Carlo experiments (mock-up data) for the different SN models and detector masses. Equation~\ref{eq:Nexp} is used to randomly determine the number of observed events $N_{\mathrm{obs}}$ from a Poissonian distribution with expectation value $\lambda = N_{\mathrm{exp}}$ for each simulated experiment. The time occurrence of each event, $t_i$, is drawn from the probability density function described in Eq.~\ref{eq:Pt}, then it is randomized with the assumed detector's time resolution of $100$~$\mu$s.

The capability of the cryogenic detector to discriminate among the representative set of models adopted in this paper is investigated through the  likelihood $\mathcal{L}$:
\begin{equation}
    \ln \mathcal{L} = N_{\mathrm{obs}}\ln N_{\mathrm{exp}} -N_{\mathrm{exp}} + \sum_{i=1}^{N_{\mathrm{obs}}}\ln p(t_i)\ .
    \label{eq:logLikTime}
\end{equation}
For one given detector mass and SN model $j$, we compute $\ln \mathcal{L}^j$ over a simulated dataset. The value of $\ln \mathcal{L}^j$ for the $j$th-model is assigned as a score of the model to the dataset. If the best scoring model for a dataset was the one used to generate that dataset, then we conclude that we are able to successfully identify the model. We stress that ours is a simple exercise to probe the detector discriminating power, given that we rely on a small set of SN models likely not representative of the whole SN population.

We generated $10^4$ different experiments for each SN model and evaluated the likelihood according to Eq.~\ref{eq:logLikTime}. The results on the RES-NOVA potential to reconstruct the SN model from the representative set of simulated experiments are summarized in Tab.~\ref{tab:discrimination}. The table shows how well each phase of the RES-NOVA detector will be able to identify each one of the six SN benchmark models (see Fig.~\ref{fig:recoils}) describing the event.
\begin{table}[]
	\centering
		\caption{Assuming the six benchmark SN models (see Fig.~\ref{fig:recoils}), the probability to identify each one of the set by means of a maximum-likelihood analysis of the expected time distribution of the neutrino signal in the three proposed RES-NOVA detectors is shown.}
\begin{tabular}{|c|r|c|c|c|}
		\cline{3-5}
		  \multicolumn{2}{c|}{~}   & \Tstrut{RN$^1$}\Bstrut & \Tstrut{RN$^2$}\Bstrut & \Tstrut{RN$^3$}\Bstrut \\
 		\hline
 		\multirow{4}{*}{\rotatebox{90}{\Tstrut Reproduced models \Bstrut}}
 		&\Tstrut$40\ M_\odot$ - slow\Bstrut	&  100\%	& 100\%	 & 100\%  \\
 		\cline{2-5}
 		  & \Tstrut$40\ M_\odot$ - fast\Bstrut &  99.7\% & 100\% & 100\%   \\
 		  \cline{2-5}
 		  & \Tstrut$9.6\ M_\odot$ - SFHo\Tstrut &  39.6\% &  92.6\% & 100\%  \\
 		  \cline{2-5}
 		 & \Tstrut$9.6\ M_\odot$ - LS220\Bstrut &  61.4\% &  93.1\% & 100\%  \\
 		  \cline{2-5}
 		 & \Tstrut$27\ M_\odot$ - SFHo\Bstrut & 52.0\% &  93.9\% & 100\% \\
 		 \cline{2-5}
 		 & \Tstrut$27\ M_\odot$ - LS220\Bstrut & 51.8\% &  98.2\% & 100\% \\
 		\hline
	\end{tabular}

 	\label{tab:discrimination}
\end{table}

Relatively to the small set of models considered in this work, the RN$^1$ detector can state without ambiguity whether the SN event led to the formation of a black hole or a neutron star, some information about the SN mass can also been inferred. On the other hand, the RN$^2$ detector can identify the progenitor mass and the nuclear EoS of the model with an accuracy $>90$\%. 
Finally, assuming the six models are the only possibilities, RN$^3$ has the outstanding potential to uniquely identify each one. While these are only six example models, this result suggests that RN$^3$ will provide excellent information about the underlying SN model.


\subsection{Estimation of the neutrino spectral features and total energy emitted in neutrinos}

The relevant astrophysical parameters that we can extract from the all-flavor neutrino signal are the average neutrino energy and the neutrino flux amplitude. In order to assess these two parameters,  we adopt the following parametrization for the time-integrated neutrino flux summed over all flavors along the lines of Eq.~\ref{eq:nuflux}~\cite{Lang:2016zhv}: 
\begin{eqnarray}
f^0(E; \langle E\rangle, \alpha_T) &=& \sum_{\beta} \int_{t_1}^{t_2} f^0_\beta(E,t)  \;dt  \nonumber = \\
 A_T \xi_T \left(\frac{E}{\langle E \rangle}\right)^{\alpha_T}&& 
 \exp\left(-\frac{(1+\alpha_T)E}{\langle E\rangle}\right)\ ,
\label{eq:parametrization}
\end{eqnarray}
where
\begin{equation*}
    \xi_T(\alpha_T, \langle E\rangle) = \frac{\left(\frac{\alpha_T+1}{\langle E\rangle}\right)^{\alpha_T+1}\langle E\rangle^{\alpha_T}}{\Gamma(\alpha_T + 1)}
\end{equation*}
is such that $\int dE f^0(E) = 1$, $A_T$ is the time-integrated neutrino flux at the detector site, and $\langle E\rangle$ is the all-flavor average energy. With such a parametrization the neutrino energy spectrum is fully described by $\langle E \rangle$, $A_T$ and $\alpha_T$. The latter is approximated with its time-average, over the $[t_1,t_2]$ time interval (corresponding to the neutronization, accretion, and cooling SN phases), and $\langle E \rangle$, $A_T$ are inferred by means of a maximum likelihood analysis, as described in the following. 

We compute  the total number of interactions $N_{\mathrm{exp}}$ (Eq.~\ref{eq:rate}) for our {\it reference model} ($27\ M_\odot$ with LS220 EoS at 10~kpc) for the three RES-NOVA detector volumes, with  $1$~keV recoil energy threshold and $200$~eV recoil energy resolution over the entire RoI: 
\begin{equation}
    N_{\mathrm{exp}} = \int \frac{d^2N}{dE_R\; dt}  \ dt \ dE_R = \int \frac{dN}{dE_R}\ dE_R\ .
    \label{eq:NexpER}
\end{equation}
The probability density function for an event to induce a recoil of energy $E_R$ is
\begin{equation}
    p(E_R) = \frac{1}{N_{\mathrm{exp}}} \frac{dN}{dE_R}\ .
    \label{eq:PER}
\end{equation}
A Monte Carlo dataset of the observable $E_R$ is generated according to the distribution of Eq.~\ref{eq:PER} and the number of generated events ($N_{\mathrm{obs}}$) is drawn from a Poisson distribution of average $N_{\mathrm{exp}}$. The parametrization in 
Eq.~\ref{eq:parametrization} is used to write the extended likelihood: $\mathcal{L}(\langle E\rangle, N_{\mathrm{exp}}(A_T);  \{E_i\}_{i=1 \dots N_{\mathrm{obs}}})$: 
\begin{equation}
\ln \mathcal{L} = N_{\mathrm{obs}} \ln(N_{\mathrm{exp}}) - N_{\mathrm{exp}} + \sum_{i = 1}^{N_{\mathrm{obs}}} \ln \tilde{p}(E_i)
\label{eq:ll}
\end{equation}
with
\begin{equation*}
\tilde{p}(E) = \frac{1}{N_{\mathrm{exp}}}N_{Pb} A_T \int dE' f^0(E'; \langle E\rangle, \alpha_T)\frac{d\sigma}{dE_R}\ .
\end{equation*}
The parameter $\alpha_T$ is fixed to its time-average over the $[t_1,t_2]$ interval.
The estimators for $A_T$ and $\langle E \rangle$ are those maximizing the likelihood in Eq.~\ref{eq:ll};  their $1\sigma$ confidence interval is obtained according to the Wilk's theorem, so that $2\ln \mathcal{L}$ drops from its maximum by the quantile of a 2D $\chi^2$ distribution evaluated at $68.26\%$. 

\begin{figure*}[htb]
    \centering 
  \includegraphics[width=0.33\textwidth]{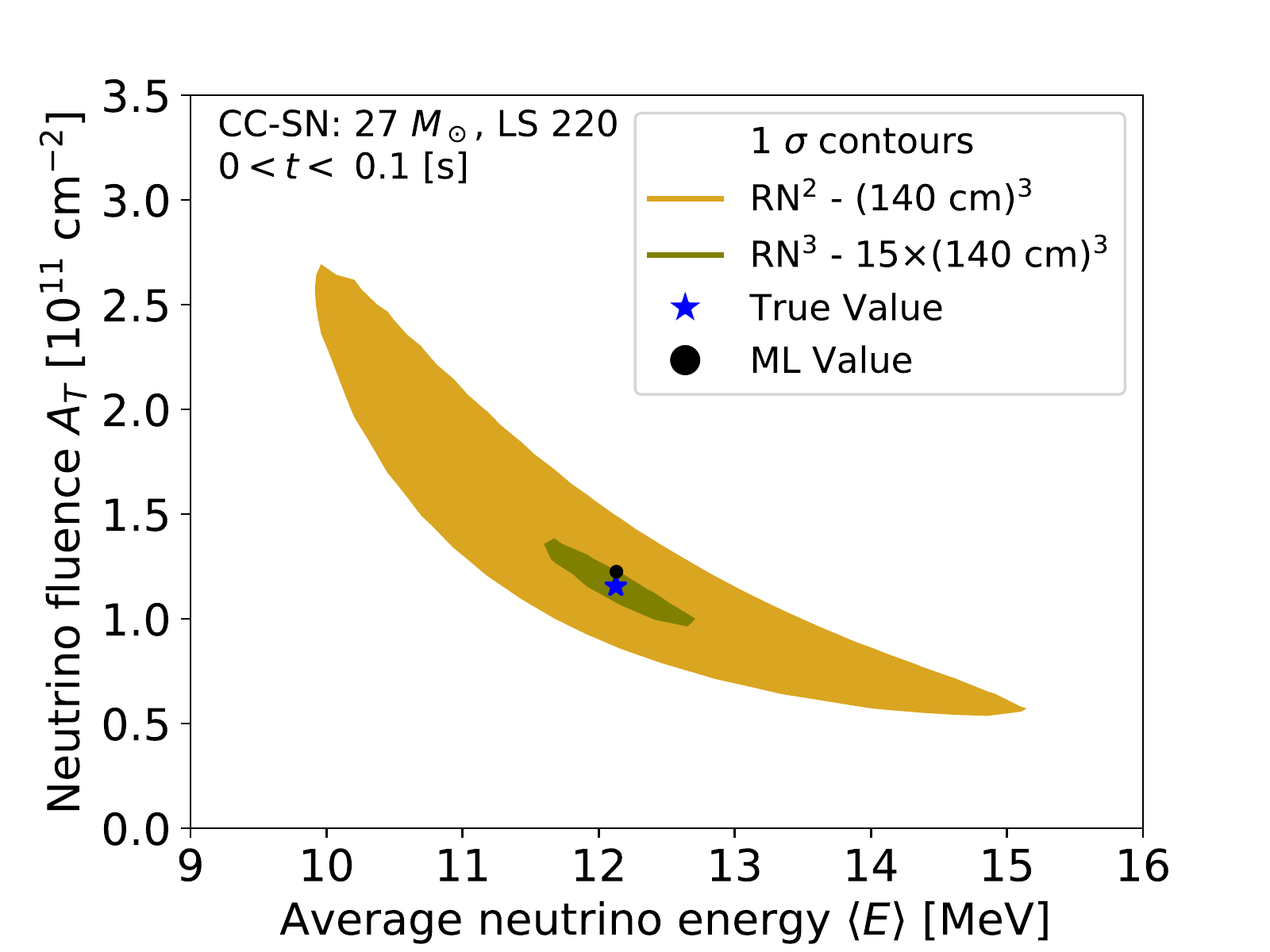}
  \includegraphics[width=0.32\textwidth]{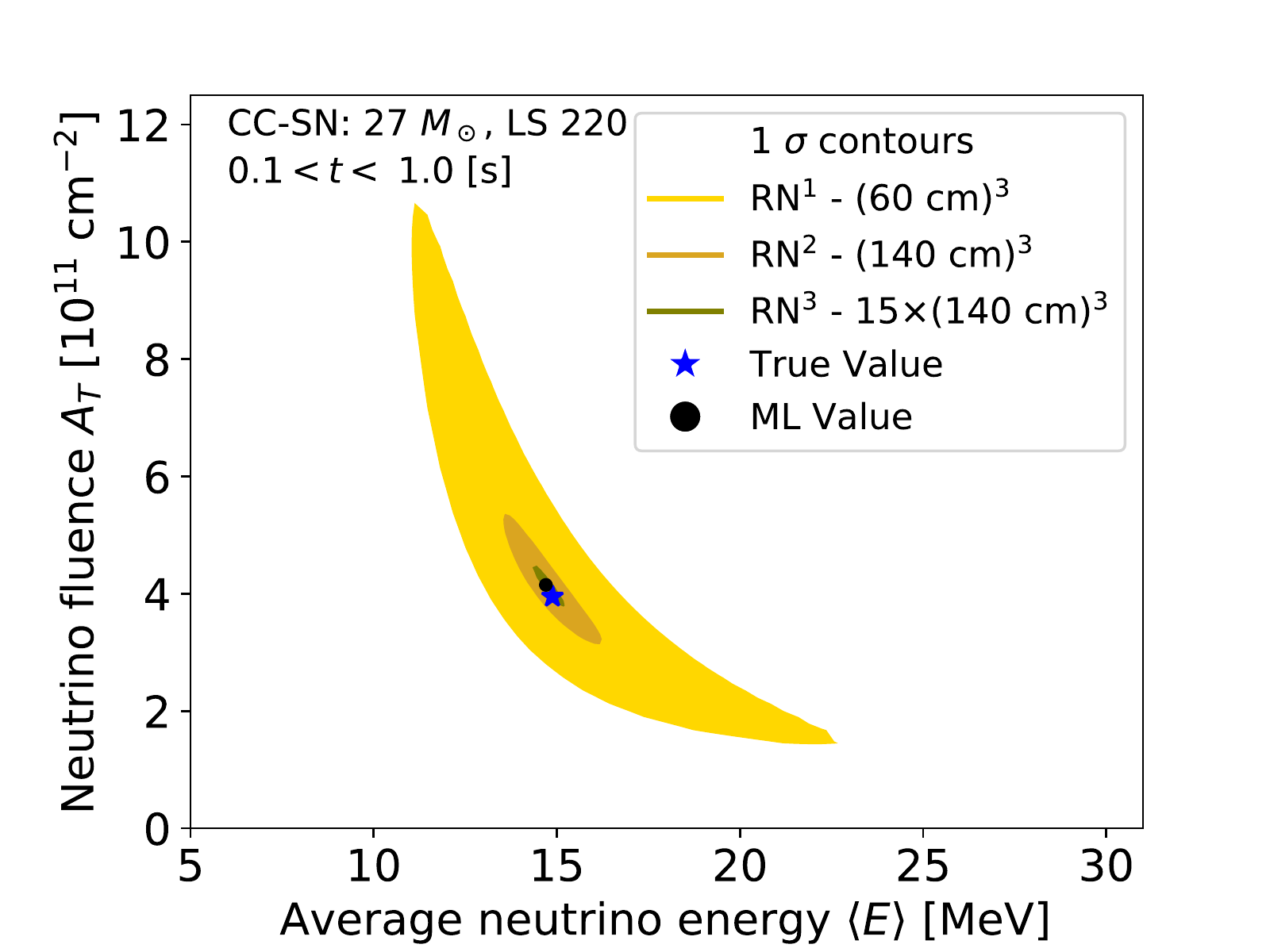}
  \includegraphics[width=0.33\textwidth]{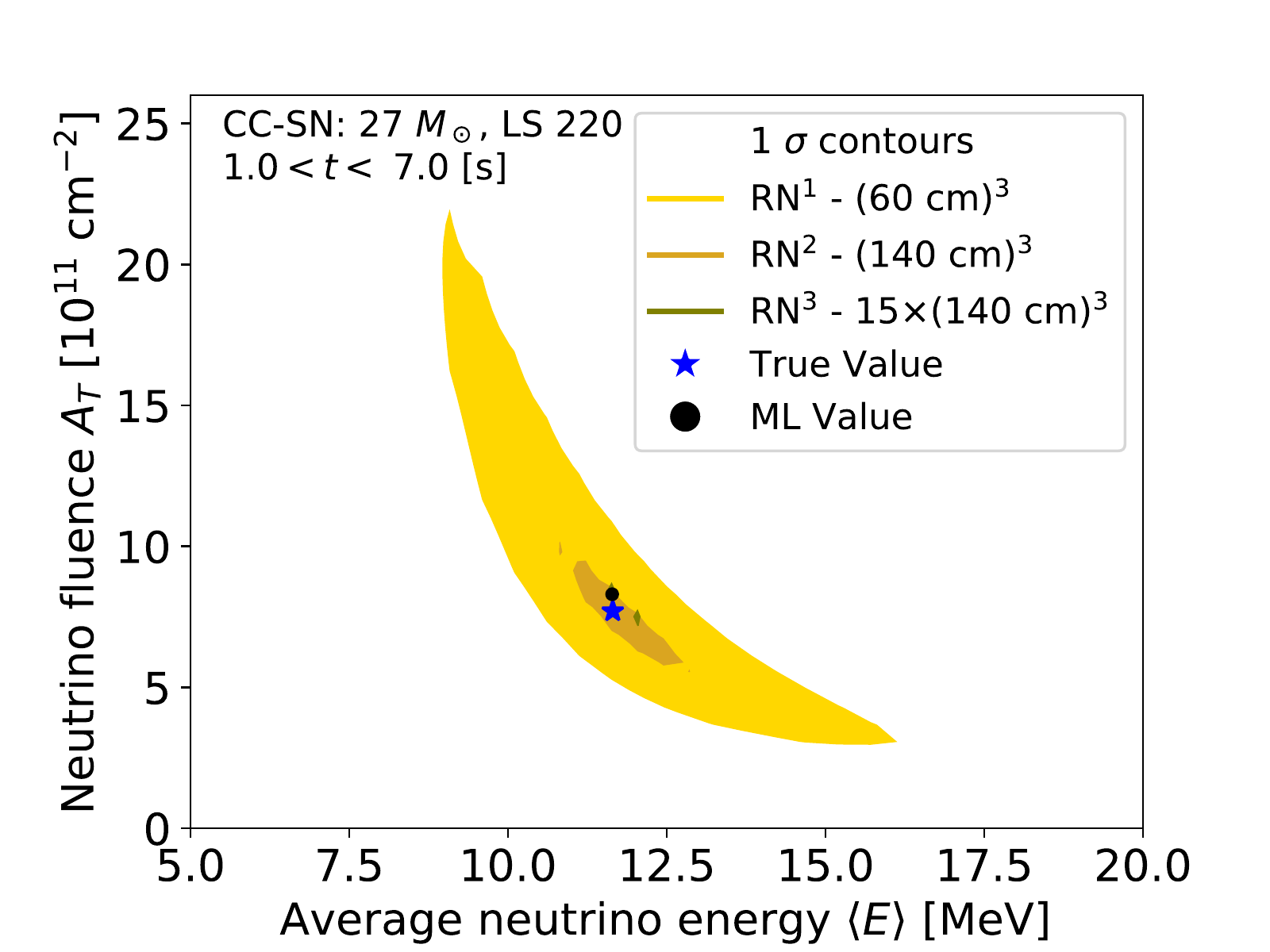}\\[0.1in]
  \includegraphics[width=0.33\textwidth]{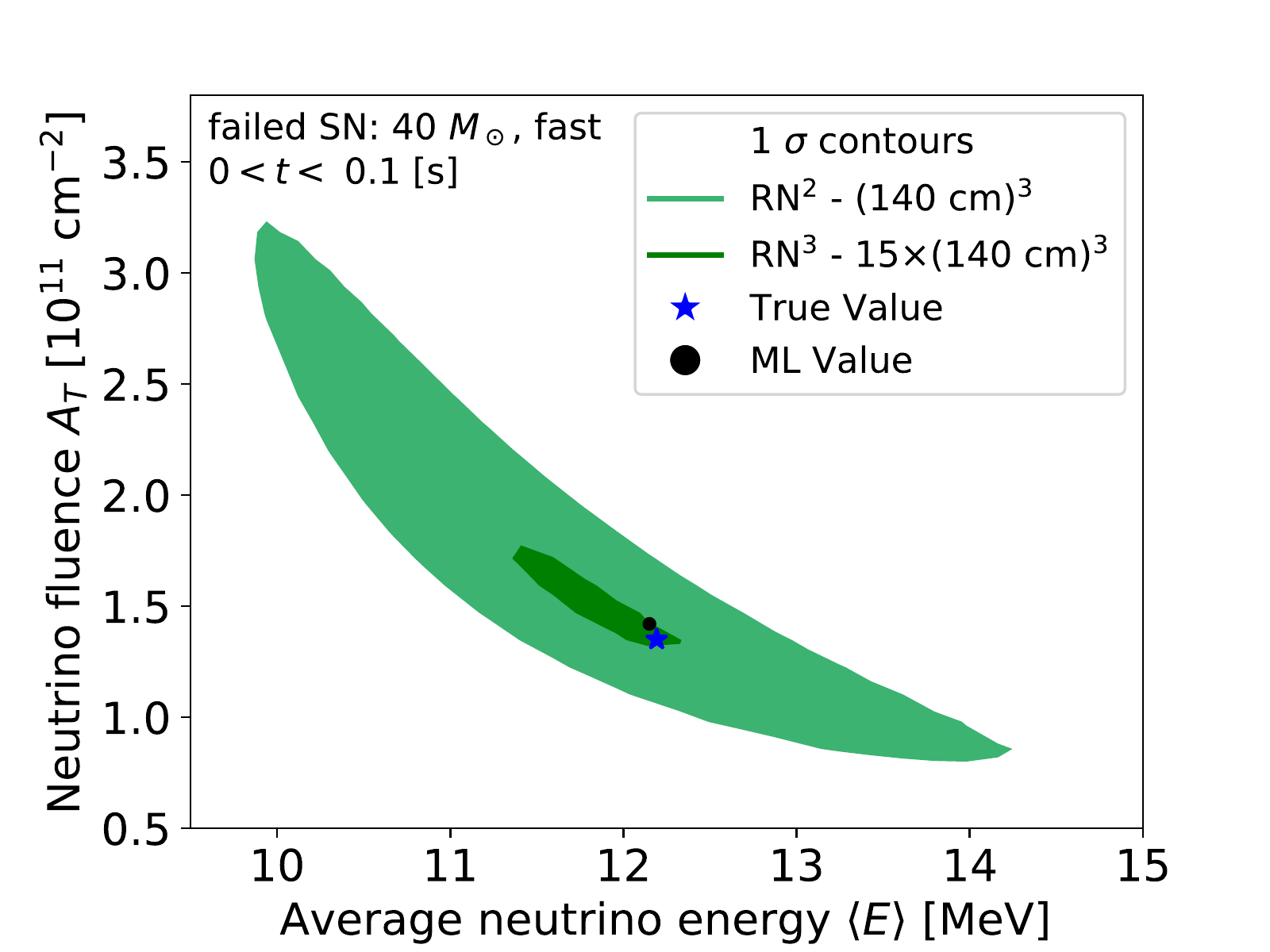}
  \includegraphics[width=0.32\textwidth]{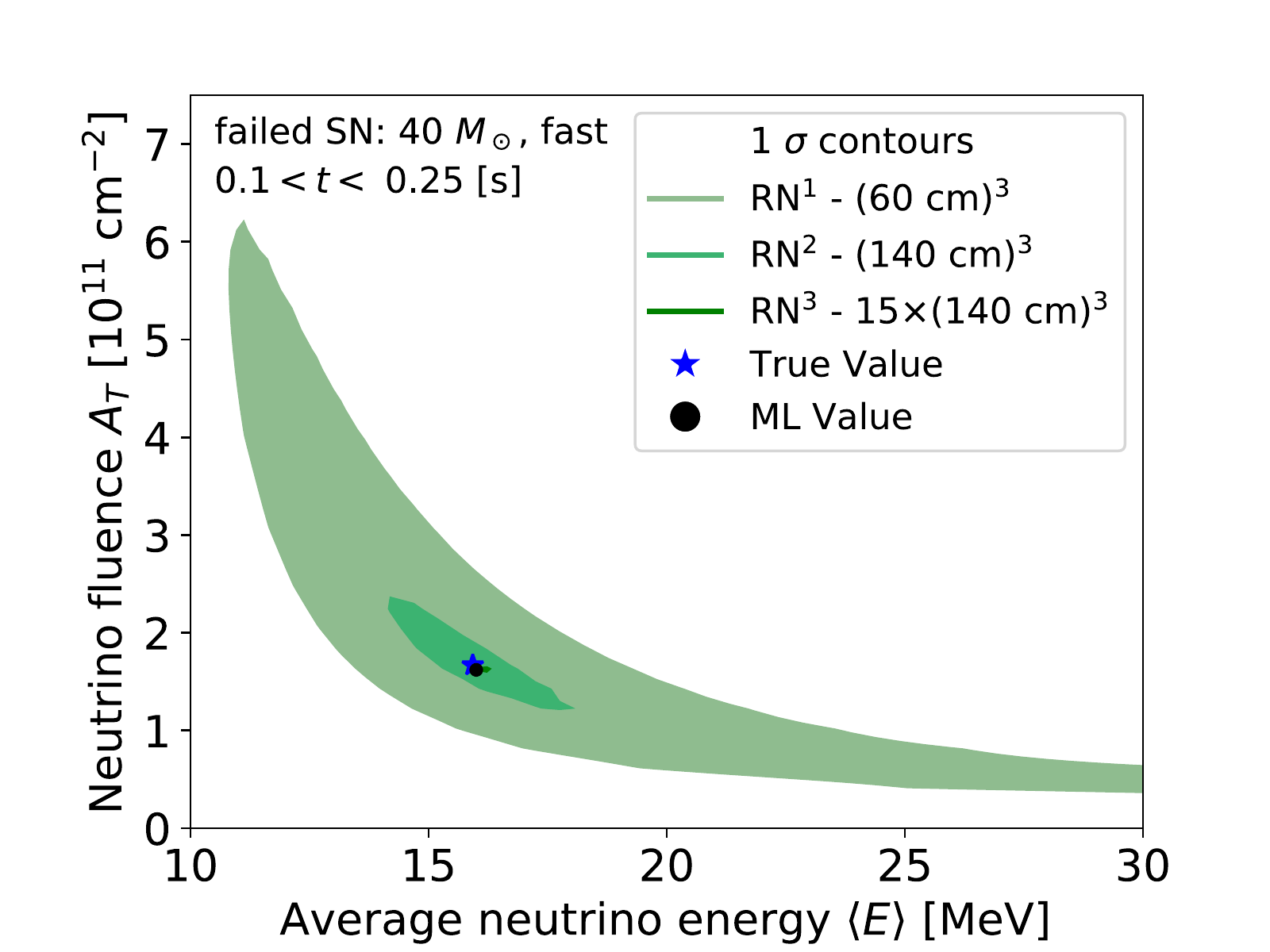}
\caption{Reconstructed average neutrino energy $\langle E \rangle$ and neutrino fluence $A_T$ at the detector site (black dot) for the $27\ M_\odot$ SN model with LS220 EoS progenitor (top panels) and for the $40\ M_\odot$ fast black hole forming collapse (bottom panels). The investigated time intervals refer to the main SN phases: neutronization ($0<t<0.1$~s), accretion  ($0.1<t< 1$~s) and cooling ($1<t<7$~s) phases, expect for the black hole forming model which is not featuring a cooling phase. The true parameter values for the different phases of the two models are shown as blue stars. The color areas are the $1\sigma$ contours for mock experiments following our maximum likelihood  analysis. The calculations are carried out for the three RES-NOVA phases: RN$^1$, RN$^2$, and RN$^3$. The small statistics  does not allow to achieve a good reconstruction of the SN parameters for the neutronization burst for both SN models in RN$^1$; hence,  only the RN$^2$ and RN$^3$ contour plots are shown. The SN parameters can be measured with excellent precision with RN$^3$.}
\label{fig:parcontours}
\end{figure*}

The projected $1\sigma$ confidence contours of  $A_T$ and $\langle E \rangle$  for the three detector volumes are shown in Fig.~\ref{fig:parcontours}. The parameter estimation is carried out for the three main phases of the neutrino emission for the $27\ M_\odot$ SN model with LS220 EoS progenitor (top panels) and for the $40\ M_\odot$ fast forming black hole collapse (bottom panels). 

The maximum likelihood reconstructed values are shown together with their respective true values for each phase of the SN burst. The small discrepancy between the reconstructed and the true values is ascribed to the numerical accuracy of the calculation.
The small number of expected events  does not allow for a reliable reconstruction of the SN parameters for the neutronization burst of the $27\ M_\odot$ SN model in RN$^1$ ($<3$~counts). For this reason only the RN$^2$ and RN$^3$ contour plots are shown. For the remaining two phases the statistics is large enough to enable an excellent parameter reconstruction with RN$^3$. Note that that the cooling phase has the longest duration among the different SN phases; as a consequence, the parametrization adopted in Eq.~\ref{eq:parametrization} is not accurate if the scanned time window is longer than $7$~s, given the large variation of the neutrino emission properties within a long time interval. In order to produce reliable results, we only consider the first $7$~s of the cooling phase, which carry most of the SN information. 

For the fast black hole forming collapse, which is not featuring a cooling phase, only two contour plots are shown. The first one refers to the neutronization phase, and also in this case the expected statistics for RN$^1$ is too low to enable a meaningful parameter reconstruction. The second contour plot shows the calculation for $[0.1,0.25]$~s. This choice is driven by the fast change in time of the flux emission parameters, namely $\langle E \rangle$ and $\alpha_T$ (see Fig.~\ref{fig:models}).  Hence, as already done for the cooling phase of the $27\ M_\odot$ SN model we select a smaller time interval.

The  reconstruction of the spectral parameters becomes extremely accurate as the detector volume increases. The SN true values for our input SN models (blue star) fall into the $1\sigma$ contour region for all the different RES-NOVA phases.  The achievable precision on the estimation of the average neutrino energy and the neutrino fluence is competitive, and in some case better, than the values reconstructed for the Xe detector in~\cite{Lang:2016zhv}, which also exploits CE$\nu$NS as main detection channel and it is therefore directly comparable. It should be stressed that our estimations rely on a very conservative assumption of the detector energy threshold of $1$~keV, while the values assumed in~\cite{Lang:2016zhv} correspond to a threshold of about $0.5$~keV. This is a significant difference in the neutrino statistics given that the spectral shape of the signal is expected to rise exponentially for lower nuclear recoil energies. Thus, RES-NOVA is able to achieve a precision on the estimation of the main SN parameters similar to the one of  direct-detection dark matter experiments, but with  less stringent constraints on the detector energy threshold. In addition, RES-NOVA has the advantage of being easily scalable to larger detector sizes.

\begin{figure}
\centering
\includegraphics[width=0.46\textwidth]{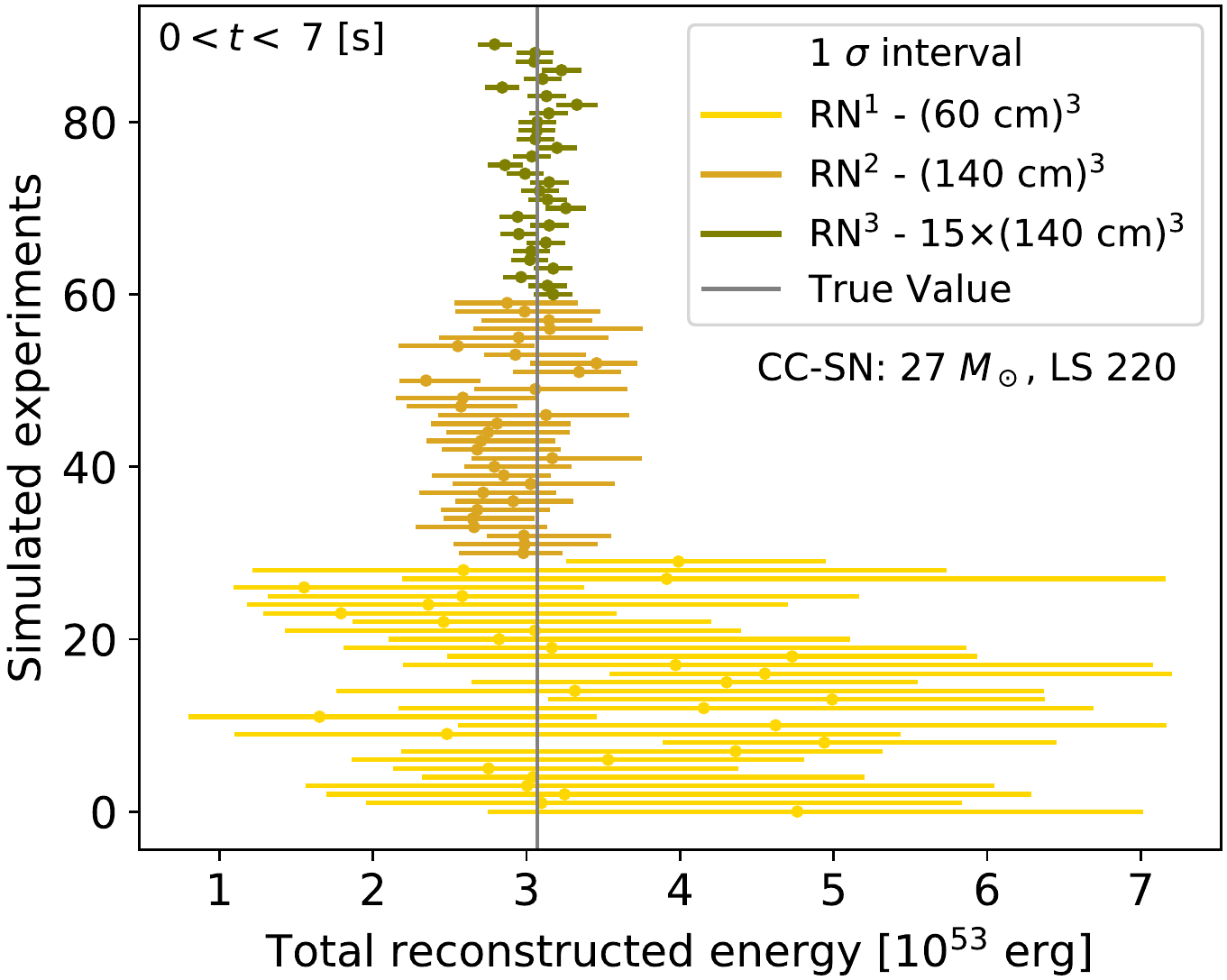}
\caption{Total reconstructed SN energy for the $27\ M_\odot$ model with LS220 EoS, for a SN at $10$~kpc.  The circles represent the maximum likelihood value while the continuous horizontal lines represent the $1\sigma$ band. The yellow (bottom), brown (middle) and green (top) bands refer to the three RES-NOVA phases, RN$^1$, RN$^2$ and RN$^3$, respectively.  The true value of the model is shown as a vertical line. The achieved precision for the three RES-NOVA detectors is: $30\%$, $8\%$, $4\%$, for RN$^1$, RN$^2$ and RN$^3$, respectively.}
\label{fig:totalenergy} 
\end{figure}

We also estimate the total energy emitted  by the SN in neutrinos:
\begin{equation}
\mathcal{E}_{\mathrm{tot}} = 4\pi d^2 A_T \langle E \rangle\ .
\label{eq:totalenergy}
\end{equation}
In order to do that, we run 30 Monte Carlo simulations, for each detector volume, to determine the precision achievable in the reconstruction of $\mathcal{E}_{\mathrm{tot}}$. The results are shown in Fig.~\ref{fig:totalenergy}, where the $1\sigma$ band for each Monte Carlo realization is plotted for  each detector volume. For this estimation, we took into account the neutrino emission interval $[0,10]$~s after the core bounce. The $1\sigma$ bands are computed propagating the uncertainty on $\mathcal{E}_{\mathrm{tot}}$ and $A_T$ in Eq.~\ref{eq:totalenergy}. The increase in the number of detected events shows a clear improvement in the precision of the  $\mathcal{E}_{\mathrm{tot}}$ measure. The shape of the contours in Fig.~\ref{fig:parcontours} directly translates into a constraint on $\mathcal{E}_{\mathrm{tot}}$ in Fig.~\ref{fig:totalenergy}. The larger and irregular contours of the RN$^1$ and RN$^2$ phases result in broader and asymmetric bounds, when compared to RN$^3$ where the very large statistics makes $\mathcal{E}_{\mathrm{tot}}$ distributed according to a normal distribution. The achieved precision for the three RES-NOVA detectors is: $30\%$, $8\%$, $4\%$, for RN$^1$, RN$^2$ and RN$^3$, respectively.

\subsection{Detection significance}
\begin{figure}[t]
\centering
\includegraphics[width=0.5\textwidth]{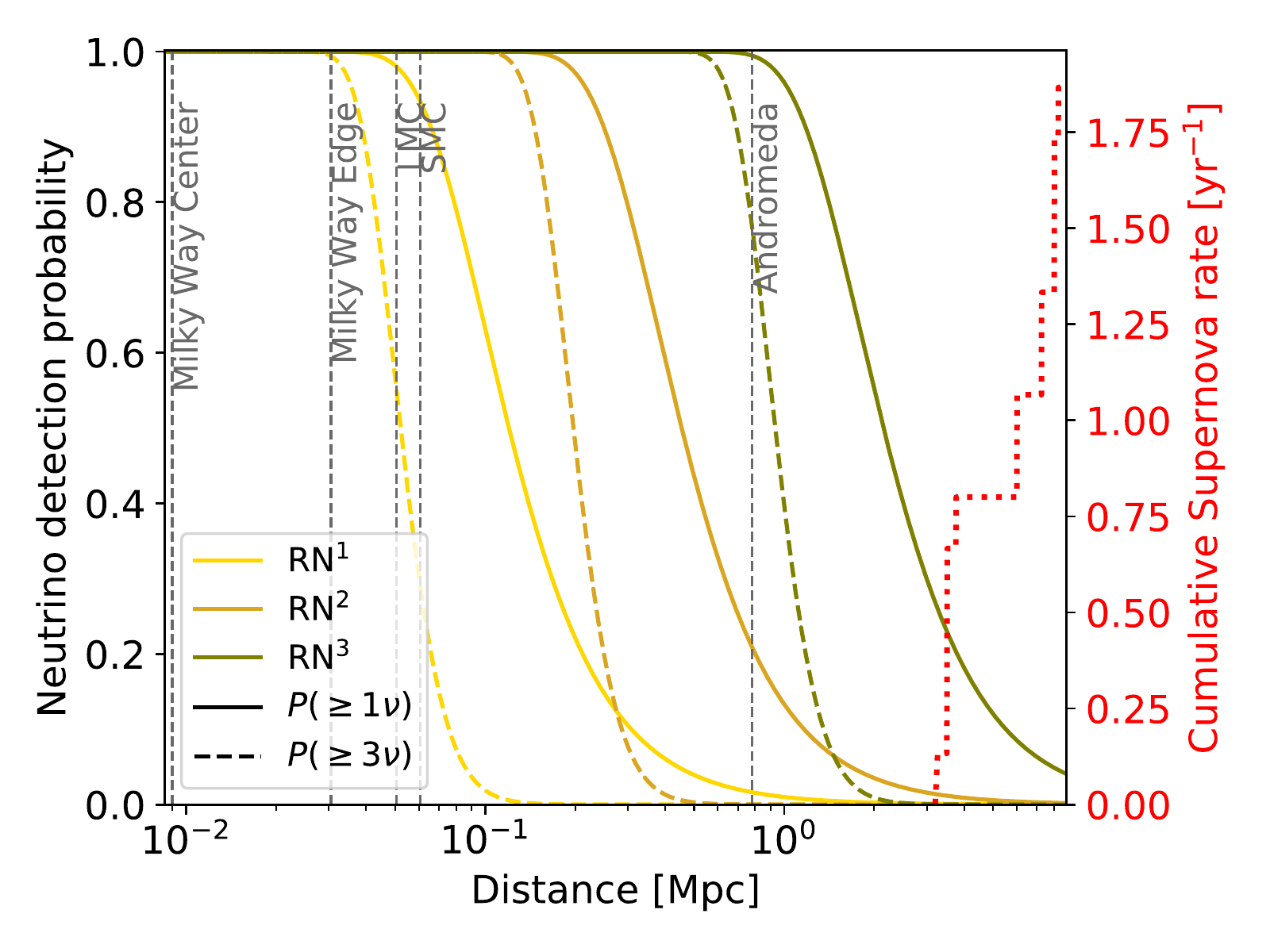}
\caption{Neutrino detection probability as a function of the SN distance for our benchmark SN model ($27\ M_\odot$ SN model with LS220 EoS). The continuous line represent the detection probability for  $\geq 1$ neutrino, while the dashed line for $\geq 3$ neutrinos. The colors refers to the three different RES-NOVA phases: RN$^1$, RN$^2$, and RN$^3$, respectively. The red dashed line shows the cumulative SN rate versus distance~\cite{Karachentsev:2004dx}. RN$^3$ will be able to detect at least one SN neutrino event every $\sim 10$~yr.}
\label{fig:poisson} 
\end{figure}

The possibility to detect SN neutrinos from neighbouring galaxies can have a strong impact on the expected rate of SN observations. In fact, the larger is the number of galaxies under investigation the higher is the probability to observe a SN event. On the other hand, the farther are the galaxies the weaker is the expected signal in the detector, as it scales as $d^{-2}$. 

Assuming that the SN neutrino search will be carried out in coincidence with other neutrino detectors and following the Poisson statistics for small signals~\cite{PhysRevD.57.3873},  we show in Fig.~\ref{fig:poisson} the probability to detect at least $1$ (continuous line) and at least $3$ SN neutrinos from a single SN explosion (using $27\ M_\odot$ SN model with LS220 EoS as benchmark model) as a function of the SN distance for the three RES-NOVA phases.
On the same plot on the right y-axis, the cumulative SN rate is also shown as from the Galaxy Catalog in~\cite{Karachentsev:2004dx}.
As pointed out in~\cite{Ando:2005ka}, the detection of a single SN neutrino, in coincidence with other SN messengers 
(e.g.,~electromagnetic radiation or gravitational waves) will enable to determine the SN bounce time with a precision of $\mathcal{O}(10~\mathrm{sec})$;  while, relying only on the electromagnetic radiation as trigger, the uncertainty would be of $\mathcal{O}(1~\mathrm{day})$.

We  infer that RN$^1$ should be able to fully explore the Milky Way, RN$^2$ can reach out to the unexplored Andromeda, and RN$^3$ may potentially  increase the expected rate of SN observations for the first time. RN$^3$ will be able to observe a SN event every $\sim3$~yr with a probability of $0.2$, thus to detect at least $1$ event every $\sim 10$~yr. Having the possibility to investigate the SN distribution over many galaxies is of utmost importance as it may shed light on the so-called SN rate problem~\cite{Horiuchi:2011zz}. Despite the promising potential of the RES-NOVA project, we stress that a thorough understanding of the detector background is needed, and the computation presented here should be considered as a benchmark study only.

\section{Diffuse supernova neutrino background}\label{sec:dsnb}

A complementary approach to the detection of a single SN burst described in the previous section is the tantalizing possibility of pushing the SN neutrino detection to cosmological scales through the DSNB~\cite{Mirizzi:2015eza,Beacom:2010kk,Lunardini:2010ab}. The DSNB has the potential to offer a glimpse on the overall SN population, opening a new epoch for extra-galactic neutrino astronomy. After introducing our theoretical DSNB model, we investigate the DSNB detectability prospects with the RES-NOVA technology.

\subsection{Theoretical inputs}
The DSNB is an isotropic and stationary flux of neutrinos and anti-neutrinos produced by past SN events, both core-collapse SNe and failed SNe. 
The DSNB summed over all six flavors is defined in the following way:
\begin{eqnarray}
    &\Phi(E)& = \sum_{\beta} \frac{c}{H_0} \int_{8\ M_\odot}^{125\ M_\odot} dM \int_0^{\infty} dz \frac{R_{\rm{SN}}(z,M)}{\sqrt{\Omega_M (1+z)^3 + \Omega_\Lambda}} \\ \nonumber &&\left[f_{\mathrm{CC-SN}} f^0_\beta[E (1+z), M] + f_{\mathrm{failed-SN}} f^0_\beta[E (1+z), M]\right]\ ,
\end{eqnarray}
where $M$ is the SN mass, $f^0_\beta(E (1+z), M)$ is the time-integrated neutrino flux  for each SN progenitor of mass $M$ (see  Eq.~\ref{eq:nuflux}), $c$ is the speed of light, $\Omega_M$ and $\Omega_\Lambda$ are the  matter and dark energy cosmic energy densities, $H_0$ is the Hubble constant, $z$ is the  redshift, and $R_{\rm{SN}}(z,M)$ is the SN rate.

Current theoretical estimations of the DSNB lack of a comprehensive knowledge of the SN explosion mechanism. Moreover, the SN distribution as a function of the redshift is still uncertain. The most relevant uncertainties affecting the expected number of DSNB events concern the neutrino mass ordering, the fraction of failed and core-collapse SNe, and the uncertainties on the SN rate~\cite{Lunardini:2012ne,Moller:2018kpn}.  

\begin{figure}[hb]
\centering
\includegraphics[width=0.48\textwidth]{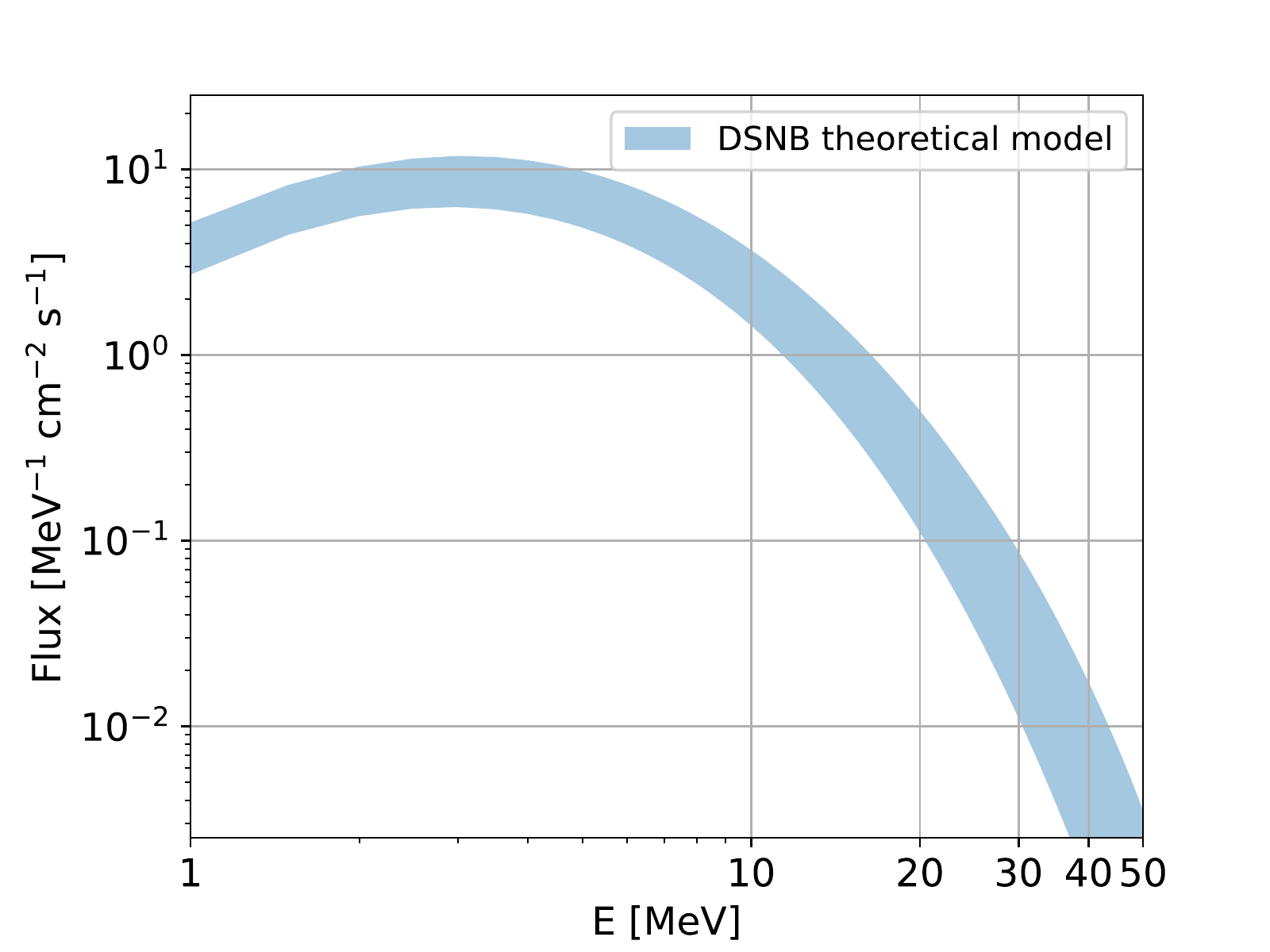}
\caption{DSNB for all six flavors as a function of the neutrino energy. The lower and upper bounds of the DSNB band were computed in Ref.~\cite{Moller:2018kpn}, see text for details}
\label{fig:DSNBtheory} 
\end{figure}

In order to explore the RES-NOVA detector response to the DSNB signal, we take into account the maximal variation of the DSNB flux computed in Ref.~\cite{Moller:2018kpn} and shown in Fig.~\ref{fig:DSNBtheory}. The lower edge of the band in Fig.~\ref{fig:DSNBtheory} was  obtained  for $f_{\mathrm{failed-SN}}=9\%$, 
$R_{\mathrm{SN}}(z=0)= 0.75 \times 10^{-4}$~Mpc$^{-3}$~yr$^{-1}$, and by adopting  the models with SFHo EoS for CC-SNe and the  $40\ M_\odot$ black hole collapse model with fast accretion as representative of the failed SN population. 
The upper edge of the DSNB in Fig.~\ref{fig:DSNBtheory} band was obtained by adopting $f_{\mathrm{failed-SN}}=0.41\%$,  
$R_{\mathrm{SN}}(z=0)= 1.75 \times 10^{-4}$~Mpc$^{-3}$~yr$^{-1}$, and  by adopting the models with  LS220 EoS for CC-SNe and the  $40\ M_\odot$ black hole collapse model with slow accretion as representative of the failed SN population. We refer the interested reader to Ref.~\cite{Moller:2018kpn} for more details on the theoretical modeling.

\subsection{Forecast of the DSNB recoil spectrum in RES-NOVA}

The DSNB has not been measured yet, nevertheless current experimental limits are very close to the theoretical upper limit. The Super-Kamiokande experiment established the most stringent limit on the $\overline{\nu}_e$ component of the DSNB flux: $>2.8 - 3.1\; {\rm cm}^{-2}\;{\rm s}^{-1}$ for neutrino energies above $17.3$~MeV~\cite{Bays:2011si}. The likelihood of detecting the DSNB, however, is dramatically enhanced by the enrichment of Super-Kamiokande with Gadolinium~\cite{Mori:2013wua} and by the upcoming JUNO~\cite{An:2015jdp}; see  Refs.~\cite{Moller:2018kpn,Priya:2017bmm} for forecasts of the DSNB detection potential with next-generation detectors.

The proposed RES-NOVA project has the unique feature of being equally sensitive to all six neutrino flavors and will be especially sensitive to the high energy tail of the DSNB. The latter carries precious information on the fraction of failed SNe on cosmological scales~\cite{Lunardini:2009ya}. Hence,  even being  a small scale detector with respect to existing Neutrino Observatories, RES-NOVA could  provide  important complementary information.

\begin{figure}
\centering
\includegraphics[width=0.48\textwidth]{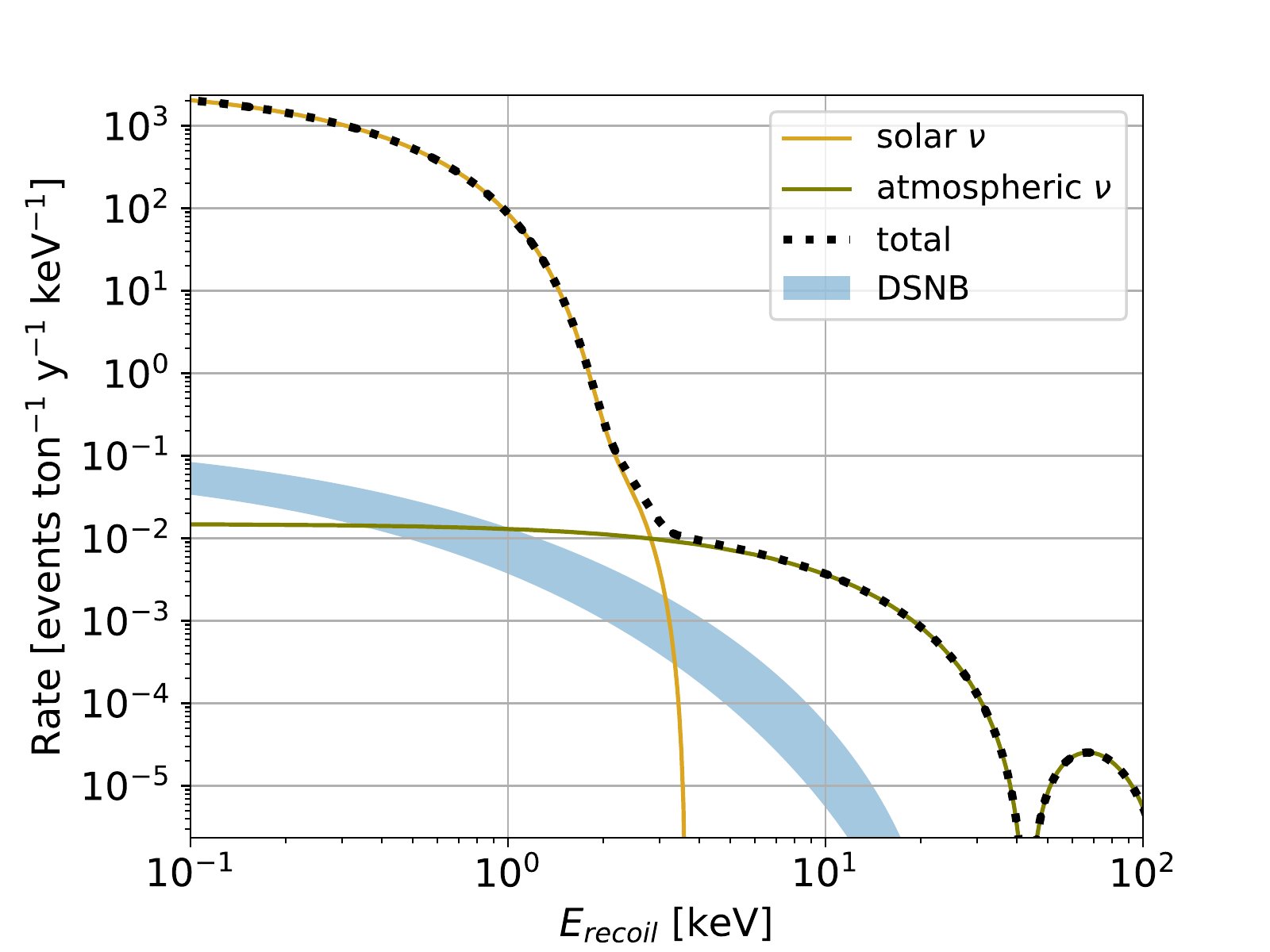}
\caption{Energy spectrum of nuclear recoils induced by neutrino interactions from different sources, per unit of detector exposure (ton $\cdot$ y). The blue band represents the maximal variation of the DSNB flux according to the calculations of~\cite{Moller:2018kpn} (see Fig.~\ref{fig:DSNBtheory}), while the continuous lines refer to the solar and atmospheric neutrino fluxes~\cite{Solar,Battistoni:2002ew}. The dotted line is the detector response to the sum of the solar, atmospheric and maximal value of the DSNB fluxes.}
\label{fig:dsnb} 
\end{figure}
Figure~\ref{fig:dsnb} shows  the spectrum of the nuclear recoils induced by neutrino interactions obtained by adopting the DSNB spectrum displayed in Fig.~\ref{fig:DSNBtheory}. The DSNB recoil spectrum is normalized for the detector exposure (ton$^{-1}$ y$^{-1}$). 
 On the same plot the expected signals from solar and atmospheric neutrino backgrounds as from~\cite{Solar,Battistoni:2002ew} are also shown. The expected DSNB signal in RES-NOVA is weak, and for this reason a through background study is mandatory. As for the natural backgrounds, the DSNB signal is overwhelmed by the solar neutrino component, mainly $^8$B, in the lower energy region (up to $\sim3$~keV). While at higher recoil energies the atmospheric background becomes dominant.

Our estimation of RES-NOVA sensitivity to the DSNB should be considered as an ideal benchmark example.  In fact, we
 here neglect the detector intrinsic background (e.g., $\beta/\gamma$'s from $^{210}$Pb in archaeological Pb) and the external neutron one (e.g., cosmic ray induced neutrons). The first background can be lowered to a negligible level by means of a particle identification technique (nuclear recoils vs. $e^-$/$\gamma$ interactions) both on pure-Pb detectors~\cite{Booth_1996} as well as on Pb-based compounds~\cite{Beeman:2012wz,Pattavina:2020ota}. The latter experimental technique (i.e., scintillating cryogenic detectors) proved a discrimination capability $>$10$^6$, depending on the light collection efficiency and on the crystal light yield at the relevant energy scale~\cite{Beeman:2012wz,Azzolini:2018dyb}. However, only direct measurements of the crystal/detector properties can provide a meaningful estimation of the discrimination power at the RoI.
The second background source can be suppressed by means of a highly efficient veto around the experimental set-up, as already done in dark matter experiments~\cite{Aprile_2014}.

If we assume a detector energy threshold of $1$~keV and collect $620$~ton $\cdot$ y of exposure, which corresponds to operate RN$^2$ for $20$~y, we expect to observe about $9$ DSNB events together with the irreducible background of solar and atmospheric neutrinos in the most optimistic scenario. Such a rate is competitive and complementary to the one expected in facilities of tens of kilotons in mass, which use water, liquid Ar, and liquid scintillator~\cite{Lunardini,Moller:2018kpn}.

\section{Conclusions}\label{sec:conclusions}
RES-NOVA is a novel detector concept, based on archaeological lead cryogenic detectors and exploiting the coherent elastic neutrino-nucleus scattering (CE$\nu$NS). The RES-NOVA research program is promisingly rich. In this paper, we explore the RES-NOVA potential to detect  supernova neutrinos.

The cryogenic technology together with the ultra-low radioactivity of archaeological lead enable an excellent detector performance in terms of energy threshold, energy resolution and background level in the region of interest. In addition, RES-NOVA can potentially achieve a high statistics detection of all neutrino flavors emitted by a supernova, thanks to the high CE$\nu$NS cross-section, which is some orders of magnitude higher than the conventional detection channels.

The innovative approach of RES-NOVA allows for a major detector miniaturization compared to  other dedicated Neutrino Observatories, without extreme detector performance (e.g., energy threshold).
Intriguingly, the modular RES-NOVA technology enables an easy detector volume scaling, without technological development, unlike all other technologies (water Cherenkov, liquid scintillator and noble-liquid time projection chambers) which run monolithic detectors. 

The scalability of the proposed technology to larger detector volumes facilitates a broad research program. In the first  phase foreseen for the detector, a thorough investigation of the entire Milky Way can be carried out; while in the second and third phases, our search can potentially reach  neighbouring galaxies.
The increase in the detector volume guarantees  large neutrino statistics, and thus to achieve a  precision up to $4\%$ on the estimation of the total  energy carried supernova neutrinos.

The original experimental approach of the RES-NOVA project promises to grant sensitivity to the diffuse supernova neutrino background with a modest detector exposure. In fact, with $620$ ton $\cdot$ y of exposure and a detector energy threshold of $1$~keV, RES-NOVA can detect about $9$ events above the irreducible background of solar and atmospheric neutrinos.

The RES-NOVA project can potentially complement currently running neutrino detectors, thanks to the high cross-section of the exploited CE$\nu$NS channel. In addition, the miniaturization of the detector and its easy scalability promise to pioneer a new generation of neutrino telescopes.

\section*{Acknowledgments}

We are grateful to Georg Raffelt and Francesco Vissani for useful discussions, and Eligio Lisi, Nirmal Raj and Anna Suliga for precious comments on the manuscript. This reasearch was supported by the Excellence Cluster ORIGINS which is funded by the Deutsche Forschungsgemeinschaft (DFG, German Research Foundation) under Germany's Excellence Strategy - EXC-2094 - 390783311, the Villum Foundation (Project No.~13164), the Danmarks Frie Forskningsfonds (Project No.~8049-00038B), the Knud H\o jgaard Foundation, and the Deutsche Forschungsgemeinschaft through Sonderforschungbereich SFB~1258 ``Neutrinos and Dark Matter in Astro- and Particle Physics'' (NDM). 

\end{document}